\documentclass{aastex}
\usepackage{emulateapj5}
\usepackage{graphics}
\usepackage{epsf}
\usepackage{color}

\def\lesssim{\mathrel{\hbox{\rlap{\hbox{\lower4pt\hbox{$\sim$}}}\hbox{$<$}}}}
\def\gtrsim{\mathrel{\hbox{\rlap{\hbox{\lower4pt\hbox{$\sim$}}}\hbox{$>$}}}}

\begin{document}

\title{The {\it K2} Light Curves and Stunted Outbursts of AC Cnc}

\author{E. M. Schlegel\altaffilmark{1,2} \& R. K. Honeycutt\altaffilmark{3}}

\altaffiltext{1}{Department of Physics and Astronomy, University of
Texas-San Antonio, San Antonio, TX 78249; eric.schlegel@utsa.edu}
\altaffiltext{2}{Vaughan Family Professor}
\altaffiltext{3}{Department of Astronomy, Indiana University, Bloomington,
     IN  47405; honey@indiana.edu}

\begin{abstract}

We describe two observations of the nova-like cataclysmic variable AC
Cnc obtained with {\it Kepler} during its revamped second mission
({\it K2}).  Using the {\it K2} 1-minute cadence mode, the data were
obtained during Campaigns 5 and 18.  Campaign 5 (C05) lasted from
${\sim}$2015 Apr 27 to ${\sim}$2015 July 10, a total of 74.8 days, and
yielded ${\sim}$106,000 measurements.  Campaign 18 (C18) lasted from
${\sim}$2018 May 13 to ${\sim}$2018 July 2, a total of 50.7 days,
yielding $\sim$72,000 measurements.  The C05 light curve reveals two
`stunted outbursts' having properties consistent with stunted bursts
observed from the ground; a stunted burst was underway during the C18
observation when it ended.  During a stunted outburst, the primary
eclipse is found to increase in depth but the residual brightness at
mid-primary-eclipse remains nearly constant.  By contrast, the
secondary eclipse retains the same depth but the brightness at
mid-secondary eclipse increases during outburst, following the orbital
variations of the out-of-eclipse light.  The eclipse ephemeris is
statistically consistent with historical ephemerides and shows only
marginal evidence for a period change.  On the basis of the {\it K2}
data, we can not confirm a previously-reported non-orbital periodicity
in AC~Cnc.

\end{abstract}


\section{Introduction}

Cataclysmic variables (CVs) are inherently variable objects, ranging
from the outbursts of novae, recurrent novae, and dwarf novae to the
flickering observed in disk systems.  Understanding the causes of the
variability requires disentangling the contributions of at least four
sources: the white dwarf, its accretion disk, the mass transfer
stream, and the red dwarf secondary.

Uninterrupted observations have led to several advances in our
understanding of CVs and white dwarfs (WDs) (e.g.,
\cite{Patterson2012, Provencal2014}).  Such observations provide a
context for variability, separating short-term variations from slower
long-term trends.

In this paper, we describe observations of the novalike CV AC~Cnc
obtained during the {\it K2} portion of the {\it Kepler} satellite
program.  The {\it K2} mission was initiated by losses of gyros, which
prevented the spacecraft from maintaining a stable pointing using the
original guiding/pointing techniques employed earlier in the mission.
{\it Kepler}, or {\it K2}, has a 95-cm diameter mirror and has two
operational modes: 1- and 30-minute observational cadences.  The
spectral response function covers ${\sim}450-850$ nm, hence it is a
`white light' telescope and detector.

AC Cnc has an orbital period of ${\sim}432.7$ min = ${\sim}0.300$ day
and an eclipse depth of ${\sim}$1.5 mag in the optical.  We now
briefly summarize studies of AC~Cnc since its discovery by
\cite{Kurochkin1980} in 1980.  \cite{Shugarov1981} argued that AC~Cnc
was an old nova.  \cite{Downes1982} detected the secondary star
through its G band and Mg $b$ absorption features, pointing to a late
G or early K star.  Based on their observations, \cite{Yamasaki1983}
inferred a large accretion disk with a steep brightness distribution.

\cite{Schlegel1984} collected time-resolved emission line spectroscopy
and obtained a mass ratio of ${\sim}1.24{\pm}0.08$ from measured
radial velocity curves of both stars.  The radial velocity curve of
the secondary was based on measurements of the G band absorption and
led to a mass of M$_2 {\sim}1.02{\pm}0.14$ M$_{\odot}$.  The radial
velocity curve of the primary was measured using several emission
lines and led to a mass M$_1 {\sim}0.82{\pm}0.13$ M$_{\odot}$.

\cite{Torbett1987} claimed a radio detection at 6 cm but
\cite{Kording2011} could not confirm the detection in a re-analysis of
the data.  \cite{Dmitrienko1995} carried out UBVRI photometry over a
4-year interval, noting that most of the variability in the 5-band
photometry likely arose from changes in the state of the accretion
disk, and discussed evidence for a reflection effect in the R and I
band photometry.

\cite{Thoroughgood2004} presented time-resolved photometry and
spectroscopy of AC Cnc. They also measured radial velocity curves of
both stars.  They obtained masses M$_1 = 0.76{\pm}0.03$ M$_{\odot}$
and M$_2 = 0.77{\pm}0.05$ M$_{\odot}$ for the secondary.  The
corresponding mass ratio is $q = 1.02{\pm}0.04$.  Their primary mass
lies within the uncertainties of the \cite{Schlegel1984} result; their
secondary mass is lower by ${\sim}$30\%; the uncertainties of that
mass from the two studies overlap at ${\sim}2{\sigma}$.  These authors
also inferred an inclination of $75^{\circ}.6{\pm}0^{\circ}.7$.

Finally, \cite{Qian2007} assembled all of the times of minima reported
in the literature as well as contributing their own measurements.
They reported detection of an orbital period change, arguing that it
was evidence of magnetic braking and a second companion.

This paper describes two pointings of the eclipsing nova-like CV AC
Cnc using the {\it Kepler} telescope in the short-cadence mode.  Here
we summarize the data and the stunted outbursts observed.

\section{Observations and Data Reduction}\label{ObsSection}

Two data sets of AC~Cnc were obtained using {\it K2}.  For the first,
{\it K2} observed the field of AC~Cnc from 2015 April 27 ${\sim}$02 UT
(JD 2457139.6006) to 2015 July 10 ${\sim}$22 UT (JD 2457214.44228) at
a one-minute cadence during Campaign 5 (C05).  The data were processed
by the {\it K2} mission center and became available in November 2015.

For the second, campaign 18 (C18) lasted from ${\sim}$2018 May 13.03
to ${\sim}$2018 July 2.92, again at one-minute cadence for AC~Cnc.
These data were also processed by the {\it K2} mission center and
became available in mid-November 2018.  A total of ${\sim}$50.7 days
${\sim}$ 169 orbits were covered.  C18 was truncated by a `hiccup' in
the fuel pump that occurred on 2018 July 2.  In response, the {\it K2}
operations team ended the campaign to preserve the data already
collected.

Hereafter, instead of referring to specific orbits with `data set 1,
orbits x to y', we will abbreviate that phrase as `orbits 1:x-y' or
`2:a-b'.  A similar construction will be used for other data
set-related quantities.

\begin{figure}
  \caption{The distribution of {\it K2} events on the pixels
    surrounding AC~Cnc for data set 1.  A light curve at each pixel
    shows the events detected by the given pixel.  The area in dark
    gray is the standard area of events extracted for the target; the
    remaining white area shows the background and possible sources of
    contamination (the black square in the upper left is an unused pixel
    required by the FITS standard \cite{Kepler2014}).
    The {\it lack} of events outside of the target area demonstrates
    that the data set is quite uncontaminated.  The plot for data set
    2 demonstrates the same conclusion and has not been included.}
  \label{pixImage}
\scalebox{0.35}{\includegraphics{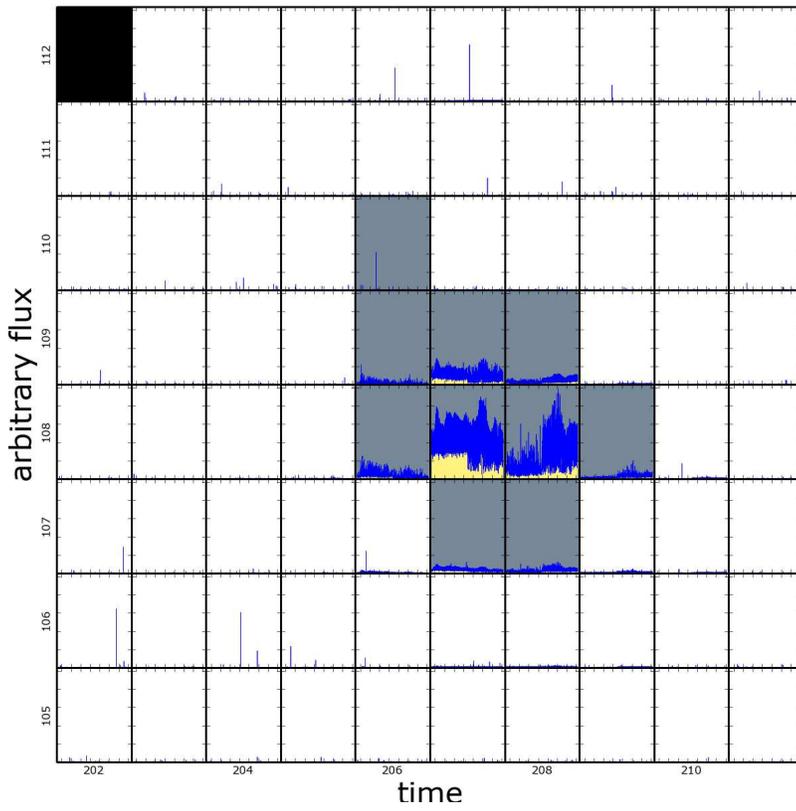}}
\end{figure}

For both data sets, we sifted through the short-cadence data using the
basic software routines of \cite{Still2012}, otherwise known as {\tt
  PyKE}, looking for artifacts as described in \cite{Kinemuchi2012} as
well as those identified by the {\it Kepler} Science Center.  We first
used the {\tt keppixseries} to plot all of the data for each pixel
(Figure~\ref{pixImage}).  This demonstrated to us that the
programmatically-defined aperture captured the vast majority of the
events from AC Cnc with minimal or zero contamination from other
stars.  That conclusion is supported by an examination of a B band
image of AC Cnc (DSS2, skyview.gsfc.nasa.gov) that shows just two
stars in the immediate region of AC Cnc: one is North by ${\sim}$5
{\it Kepler} pixels while the other is South by ${\sim}$5-6 pixels.
Those fortuitous stellar positions leave the light from AC~Cnc
uncontaminated to a high degree.  Consequently, we did not apply any
filtering to the data at this stage.

Short gaps occur in the data, usually lasting $<$0.005 in phase for
data set 1, but longer for data set 2.  Generally, the short gaps are
associated with cosmic ray hits, data dumps, and attitude corrections.
During C18, there are the usual cosmic ray hit-generated gaps.  But
there are also a number of small gaps as the spacecraft dropped into
`coarse pointing' rather than fine pointing, then re-gained fine
pointing control.  The longest gap occurs at orbits 2:46-47, when
little data are collected during a gap lasting nearly a full orbit.
We do not make any attempt to fill in gaps.

The short-cadence data then contains 1:109,890 and 2:73,728 individual
points.  We filtered both data sets to eliminate observations for
which the SAP\_QUALITY flag was non-zero, as recommended by the {\it
  K2} team.  Cosmic ray hits and coarse pointing times comprised the
bulk of the non-zero SAP\_QUALITY flags.  After filtering, the data
sets had a total of 1:106,703 individual points covering the 74.8 days
of the pointing and 2:72,549 points covering 50.7 days.  Given that
the period of AC~Cnc is ${\approx}$0.3 days, we consequently expect
${\approx}$1:240-250 and ${\approx}$2:175 successive eclipses; we
count 1:246 and 2:169 eclipses.

\subsection{Observational Precision}\label{precision}

The photometric precision is limited by spacecraft jitter,
demonstrated to be only a few percent larger than during the {\it
  Kepler} core mission, and the much larger solar wind-induced drift,
which moves the source across pixels of varying sensitivity
\citep{Howell2014}.  In that paper, the photometric accuracy is
presented as median 6-hour precision in parts per million (ppm),
showing that the {\it K2} mission is about a factor of ${\approx}$4
worse than for the {\it Kepler} core mission -- instead of ${\sim}$20
ppm at mag 12, {\it K2} delivers ${\approx}$80 ppm.

However, the median 6-hour precision is not helpful given that our
AC~Cnc data sets have 1-minute cadence.  A more direct measure of
precision is available in \cite{Howell2014} who re-observed the
transiting exoplanet WASP-28b, demonstrating 0.16\% precision.  If we
then degrade that precision to a mag 13.5 object, we expect
observations of AC~Cnc at 1-minute cadence to possess ${\approx}$0.5\%
precision.  Consequently, for an observation yielding 10,000 counts,
we expect $\lesssim$50 counts of uncertainty.  This is a factor of
${\approx}$1.5 {\it above} the uncertainties included in the {\it K2}
data file.  The variations in the light curve we describe are all
substantially larger than these uncertainties.

As an aid to the reader, while we have not converted the data points to
magnitudes, we could do so using the relation from the {\it Kepler}
team between {\it K2} counts and magnitude:

$$f_{kep} / 1.74{\times}10^5 = 10^{-0.4}(K_{p2} - 12)$$

{\noindent}or

$$K_{p2} (mag) = -2.5 log( f_{kep} ) + 25.10137.$$

{\noindent}where $f_{kep}$ are the {\it K2} counts.

\subsection{The Ephemeris}\label{Ephem}

We define the minimum of each eclipse by fitting, for each eclipse,
times ${\approx}{\pm}$15 min about the approximate eclipse minimum
with a second-order polynomial.  These times of minima are then fit
with the standard ephemeris equation to yield the period.  Our initial
ephemeris is then

 $$BJD = 2457139.32155({\pm}24) + 0^d.300482312({\pm}15) {\times} E$$

{\noindent}where the uncertainties are in the last two digits.

The {\it Kepler} program returns times as BJD while all previous
observations of AC~Cnc are HJD (q.v. \cite{Qian2007} for the most
up-to-date list of eclipse minima).  As noted by \cite{Eastman2010},
the difference between HJD and BJD is a semi-sinusoid with an
amplitude up to ${\pm}$4 seconds ${\sim}0^d.46{\times}10^{-4}$ for an
object in the ecliptic plane.  The sinusoid is primarily caused by the
changing positions of Jupiter and Saturn.  AC~Cnc's ecliptic latitude
is ${\sim}$-5$^{\circ}$, which has negligible effect on that
sinusoidal amplitude for our purposes.

While differences of a few seconds are small with respect to the
${\sim}0^d.005$ (O~-~C) trend displayed in \cite{Qian2007}, the
differences are {\it not} small relative to the uncertainties of the
times of minima.  We presume the period and $\dot{P}$ terms are
relatively independent of the time references used.  We therefore fit
the data three times: firstly, we fit the two {\it K2} datas sets
independently to verify that we had the correct cycle count and to
check for a $\dot{P}$ term over the ${\sim}$3000 orbits between C05
and C18.  Secondly, we fit the combined {\it K2} data and the HJD data
sets, for which we permitted the ephemeris start times and the
$\dot{P}$ terms to be independent, but set the orbital period to be
simultaneously fit for the HJD and BJD data sets.  Finally, we set the
$\dot{P}$ term equal between the HJD and BJD data sets and fit for the
ephemeris start time and period.

\begin{figure}
 \caption{Contours for times of minimum: (a) historical period (HJD)
    vs K2-only period (BJD); and (b) $\dot{P}_{historical}$ vs
    $\dot{P}_{K2 only}$, both from the unconstrained fit; and (c) historical
    period (HJD) vs K2 period (BJD) with $\dot{P}$ equal for the
    two data sets.  The lines in (c) indicate equal periods relative
    to the best-fit solution.  Note that the horizontal and vertical scales
    differ in scale by 10x in (a).}
\label{TimeFigs}
\scalebox{0.5}{\includegraphics{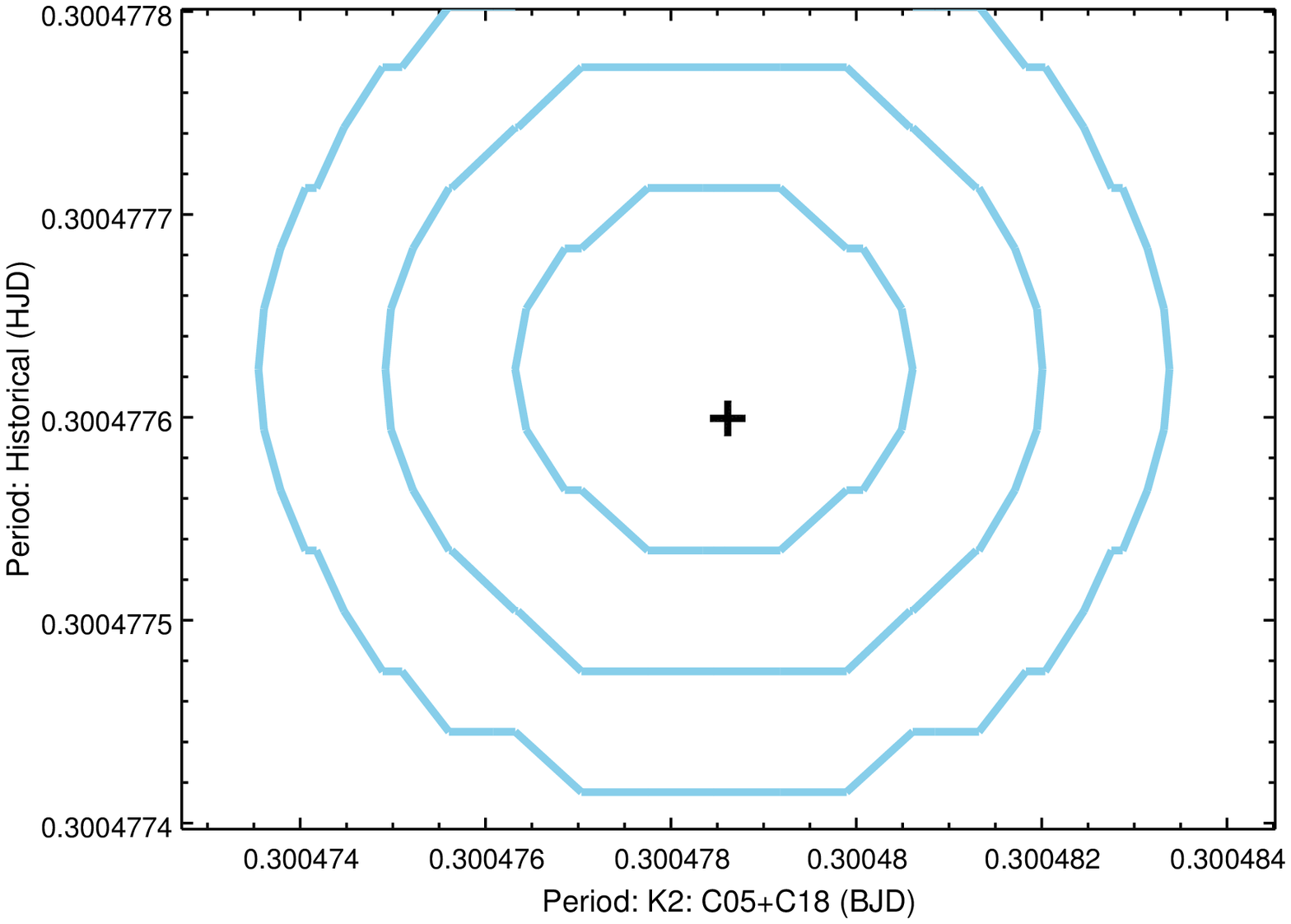}}
\scalebox{0.5}{\includegraphics{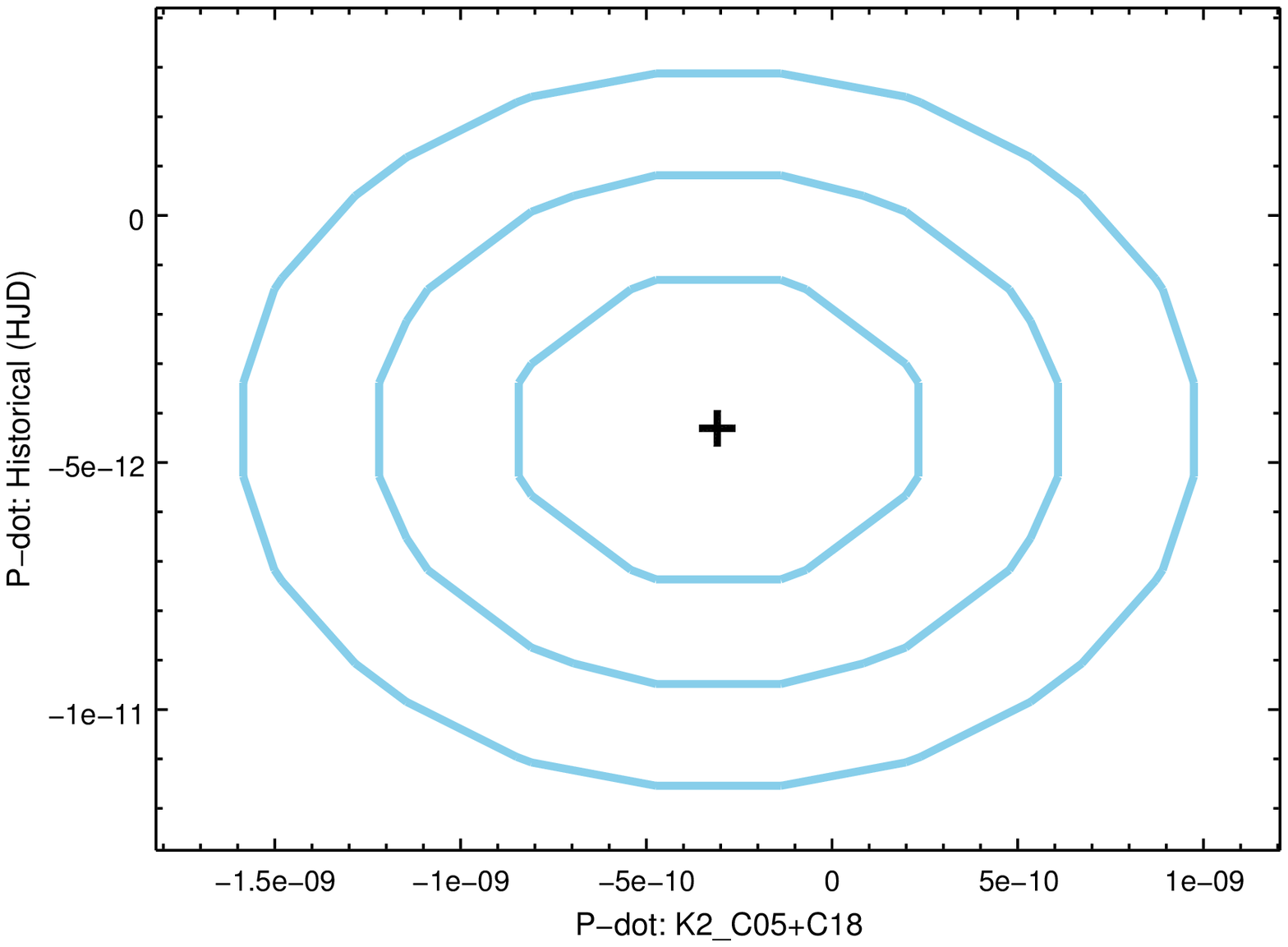}}
\scalebox{0.5}{\includegraphics{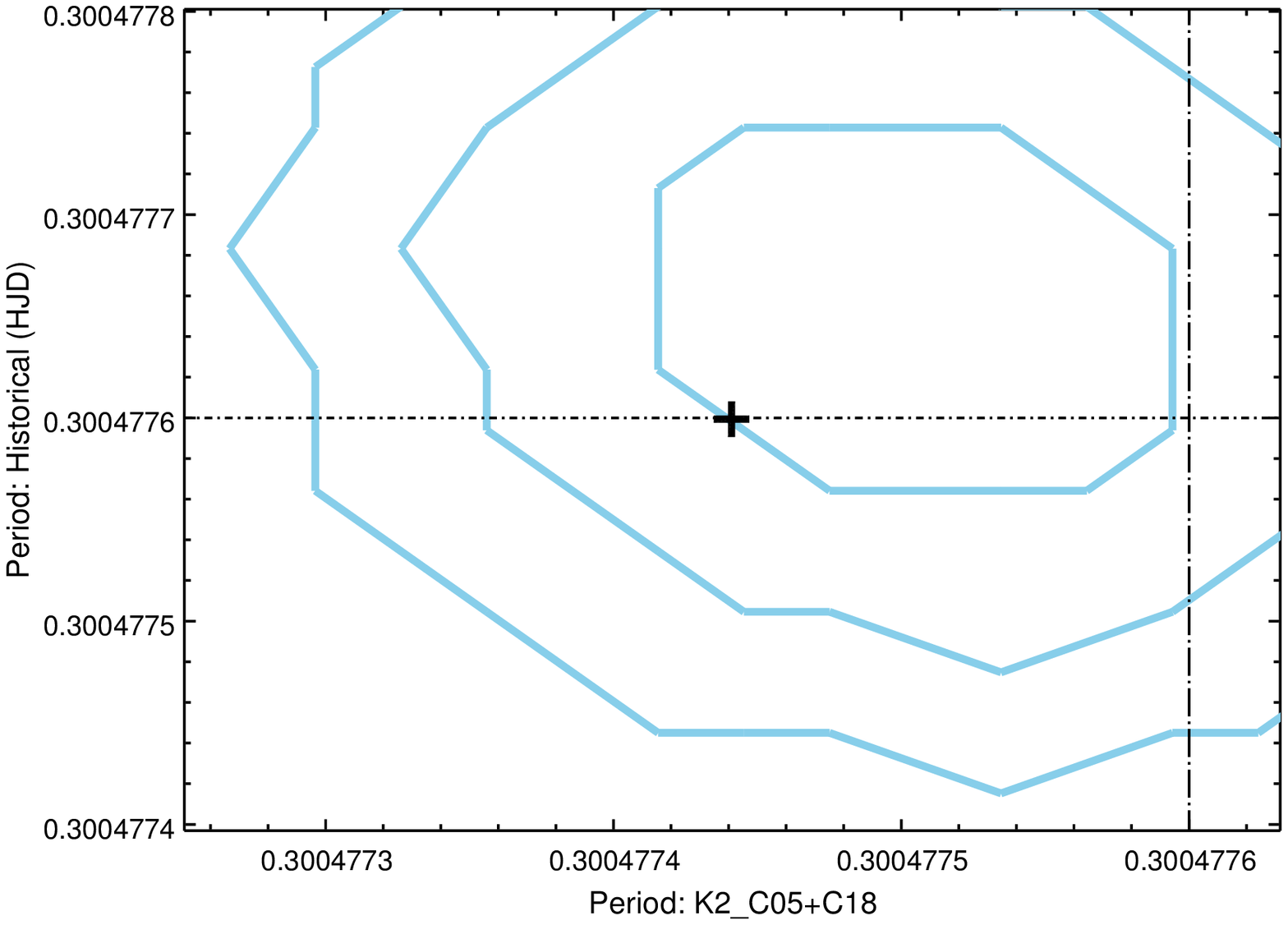}}
\end{figure}

The results are shown in Table~\ref{TimeFit} and in
Figure~\ref{TimeFigs}.  The completely unconstrained fit is the
critical one.  Note that the periods are not statistically different.
However, the $\dot{P}$ term for the BJD data is consistent with zero.
For the historical data, the consistency with zero lies just inside
the 90\% contour, suggesting at this point at best marginal evidence
for a changing period.  Note, however, the magnitudes of the two
$\dot{P}$ terms: ${\sim}4{\times}10^{-10}$ (BJD) vs
${\sim}2{\times}10^{-12}$ (HJD).

\begin{table*}
\begin{center}
\caption{Fits for Eclipse Ephemeris\label{TimeFit}}
\begin{tabular}{|l|rr|rr|rr|}
\hline
     & \multicolumn{2}{|c|}{$\dot{P}$-constrained} & \multicolumn{2}{|c|}{Period Constrained}   & \multicolumn{2}{|c|}{All Unconstrained} \\
 Quantity    & Value         & Uncertainty  &   Value       & Uncertainty & Value  & Uncertainty \\  \hline
%             &               &              &               &             &      \\
  & \multicolumn{2}{|c|}{Historical times (HJD)} & \multicolumn{2}{|c|}{Historical times (HJD)} & \multicolumn{2}{|c|}{Historical times (HJD)} \\
 Time-zero$^a$ & 44290.30768 & $\pm$3.1e-4  & 44290.30768 & $\pm$3.1e-4  & 44290.30769 & $\pm$3.1e-4  \\
 Period        & 0.300477599 & $\pm$5.0e-8  & 0.300477600 & $\pm$5.0e-8  & 0.300477599 & $\pm$5.0e-8  \\
 $\dot{P}$     & -4.3e-12    & $\pm$2.1e-12 & -4.3e-12    & $\pm$2.1e-12 & -4.3e-12    & $\pm$2.1e-12 \\
               &             &              &             &              &             &    \\
  & \multicolumn{2}{|c|}{{\it K2} C05+C18 (BJD)} & \multicolumn{2}{|c}{{\it K2} C05+C18 (BJD)} & \multicolumn{2}{|c|}{{\it K2} C05+C18 (BJD)}      \\
 Time-zero$^a$ & 57139.32172 & $\pm$1.2e-4  & 57139.32170 & $\pm$1.2e-4  & 57139.32156 & $\pm$2.2e-4 \\
 Period        & 0.300477441 & $\pm$4.7e-8  &  $\cdots$   &  $\cdots$    & 0.30047862  & $\pm$1.5e-6 \\
 $\dot{P}$     &  $\cdots$   &  $\cdots$    & -4.5e-11    & $\pm$1.7e-11 & -3.07e-10   & $\pm$3.8e-10 \\
\hline
\end{tabular}

$^a$Time zero is JD - 2,400,000.  Uncertainties are 90\% confidence limits.
\end{center}
\end{table*}

We acknowledge that our `solution' is {\it not} the best way to
address this problem.  The correct approach would convert the original
JD times to BJD times.  That study lies outside the scope of this
paper because we lack the original JD times.  Our approach
demonstrates the increasing necessity of converting {\it all} times to
a consistent reference frame, as noted by \cite{Eastman2010}, because
the precision is necessary to assess the possibility of changing
periods.  Given the results in Table~\ref{TimeFit} and the contours in
Figure~\ref{TimeFigs}, it is possible a $\dot{P}$ term exists but, at
this point, is not supported by the data.  This conclusion does not
affect any of the remainder of this paper.

\section{The Light Curves}

\subsection{Overall Light Curve and Primary Eclipse Behaviors}

Figure~\ref{CompleteLC} shows the complete, filtered data sets at
1-minute cadence using the {\tt K2} flux values (a, b = set 1, 2).

The points appearing near a flux level of ${\approx}$13,000 are the
bottoms of the primary eclipse.  The denser band of points falling
${\approx}$5000 counts below the upper-most points of the light curves
at all phases represent the bottoms of the secondary eclipses.  Two
arrows in each figure point to these bottom points, respectively.

That the primary eclipse drops the light to a nearly constant value
tells us that the brightness variations are not visible during the
eclipse.  That behavior also shows that uneclipsed accretion
structures (i.e., the outer portions of the disk and the stream) do
not vary much in location, shape, or brightness from eclipse to
eclipse.

In contrast, the light remaining at the bottom of the secondary
eclipse is highly variable on time scales of days to weeks, and these
changes track the changes outside eclipses.  This implies that the
day-week variability in AC Cnc arises in some combination of the inner
portions of the accretion disk, the inner accretion stream, or the
inner hemisphere of the secondary star.  This is similar to the
conclusions of \cite{Dmitrienko1995} for the location(s) of
variablility in AC Cnc, based on the behavior outside and in the
middle of eclipses.

Table~\ref{HalfWidths} lists, and Figure~\ref{EclipWidth} shows, the
primary eclipse half-widths at half-maximum across both {\it K2} data
sets.  These widths are nearly constant at ${\sim}$0.094 in phase --
the median and mean values differ slightly: median: 1:0.0952 vs mean:
1:0.0932 and 2:0.0943 vs 2:0.0929.  Within the uncertainties, however,
these are identical.  The half-width start phases have a mean and
median of 1:0.051 and 1:0.052 and 2:0.052 and 2:0.052, respectively.
The corresponding end phase values are (mean, median) 1:0.042 and
1:0.043 and 2:0.041 and 2:0.042, respectively.  Within the
uncertainties, the two sets of mean and median values are also
identical.  The difference between the start and end phase may
indicate a slightly asymmetric brightness distribution within the
disk.

\begin{table*}
  \begin{center}
    \caption{Median Eclipse Half-widths\label{HalfWidths}}
    \begin{tabular}{lrrr|rrr}
      \hline
               & \multicolumn{3}{c|}{Data Set 1} & \multicolumn{3}{|c}{Data Set 2} \\
      Eclipse  & Phase In$^a$         & Phase Out$^a$       & Width               &  Phase In$^a$        & Phase Out$^a$        & Width \\
      Primary  & -0.0525${\pm}$0.0005 & 0.0430${\pm}$0.0004 & 0.0952${\pm}$0.0013 & -0.0525${\pm}$0.0008 & 0.0418${\pm}$0.0004 & 0.0943${\pm}$0.0017 \\
     Secondary & -0.0605${\pm}$0.0018 & 0.0612${\pm}$0.0022 & 0.1217${\pm}$0.0057 & -0.0650${\pm}$0.0027 & 0.0601${\pm}$0.0028 & 0.1247${\pm}$0.0078 \\
     \hline
    \end{tabular}

    $^a$All phases listed in the table are relative to phase 0.5.
  \end{center}
\end{table*}

We note that there are essentially zero pre-eclipse `bright
spots' as sometimes observed in nova-like CVs (e.g.,
\cite{Warren2006,Dhillon2013}).  Of particular note is the lack of
pre-eclipse brightening before the rise of or during the bursts.

Finally, comparing the two data sets leads to two different
descriptions: data set 1 shows a nearly-continuous variability in
total light while data set 2 shows nearly constant total light.
Interestingly, data set 2 has a mean out-of-eclipse level of
${\sim}$28K counts while data set 1's mean is ${\sim}$23K counts.  We
can not make too much of this observation as we lack a sufficiently
large data set to place those numbers in context.

\subsection{Stunted Outbursts}

Both data sets exhibit what appear to be stunted outbursts, as
indicated by the labels.  We compare their properties to the
literature on stunted bursts in greater detail in the next section.

\begin{figure}
  \caption{The complete {\it K2} light curve of AC~Cnc.  Note the
    bottoms of the primary and secondary eclipses in each light curve
    as delineated by the colored arrows.  Also note the two stunted
    outbursts in data set 1 and the initial phases of burst 3 in data set
    2.  Phases are assigned based on the ephemeris listed in
    \S\ref{ObsSection}.  The colored boxes are extracted and shown in
    Figure~\ref{ExpLC}.  The smooth line labeled `Background'
    indicates the background counts boosted by a factor of 2.  The
    green arrows (data set 1) and tilted line (data set 2) indicate the
    responses of the primary minima to the outbursts.
    (a) Data Set 1}
  \label{CompleteLC}
\scalebox{0.5}{\rotatebox{-90}{\includegraphics{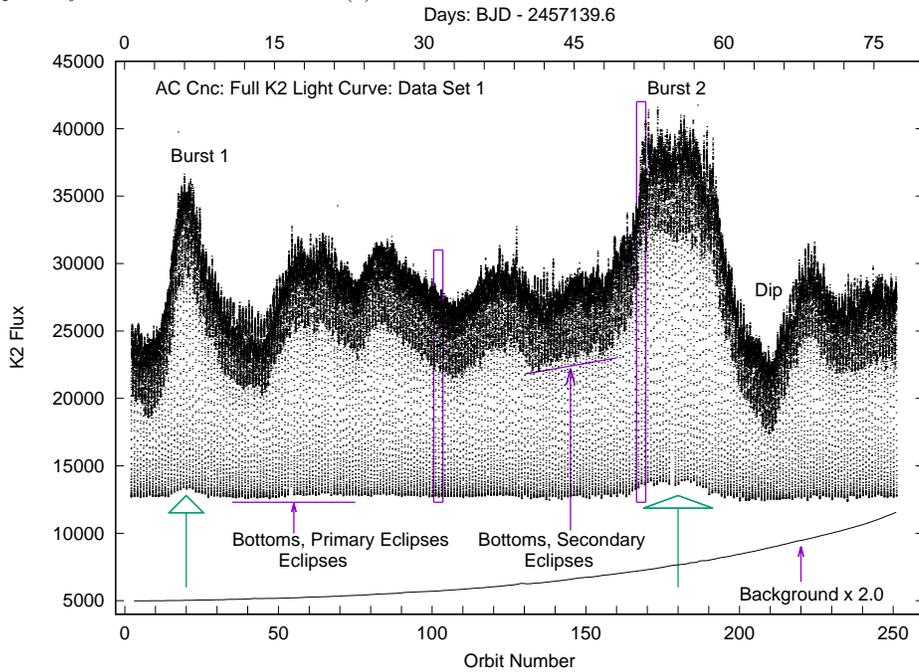}}}
\end{figure}

Figure~\ref{ExpLC} shows expansions of two sections of the data set
1 light curve at orbits away from an outburst (a) and on the rise of
the larger burst (b).  The expanded regions are indicated by the boxes
in Figure~\ref{CompleteLC}(a).  These are roughly representative of
the range of behavior in both data sets.

We note that burst 2 appears to distort the eclipse ingress,
shortening it from phase ${\sim}0.95$ to ${\sim}0.97-0.98$ as is
visible in Fig.~\ref{EclipWidth}(a).  Curiously, there is little
apparent effect from burst 1 on the light curve.  Burst 3, while not
complete, also appears to decrease the ingress time near the peak of
the burst.

Figure~\ref{PhzdLC}(a) shows a fully-phased light curve of data set
1 and \ref{PhzdLC}(b) of data set 2.  For data set 1, the heavy
band in the center of the cloud of points is essentially the `mean'
light curve.  The complete range for non-eclipse phases is
${\approx}$22K counts to ${\approx}$38K counts which corresponds to
${\sim}$0.6 mag.  Of that 0.6 mag, ${\approx}$0.4 represents the burst
and the remaining ${\approx}$0.2 mag represents variations of the mean
light curve.  For data set 2, a similar description holds.

Excursions below the mean are largely due to the dip that follows
Burst 2 (which can be seen in Figure~\ref{CompleteLC}(a) at orbits
${\sim}1:200-215$).  Such outburst/dip pairs are occasionally seen in
other NL CVs having stunted outbursts, as noted in
\cite{Honeycutt1998} and \cite{Honeycutt2001}.  Dips following
stunted outbursts have also been documented in {\it Kepler} photometry
of KIC 9406652 \citep{Gies2013}, V523 Lyr \citep{Mason2016}, and 
KIC 9202990 \citep{Ramsay2016}.

Figure~\ref{PhzdLC} is useful to demonstrate the excursions from a
mean behavior.  But plotting all 246+169 light curves makes it
impossible to follow a specific light curve.  In
Figure~\ref{plotlets}, we plot the light curves in blocks of ten.  We
immediately are able to pick out the orbits during which the stunted
outbursts originated (orbits ${\approx}$1:11-30 for burst 1;
${\approx}$1:161-190 for burst 2; and orbits ${\approx}$2:151-160 for
burst 3).  We also can identify the orbits during which the dip
occurred (orbits ${\approx}$1:201-220).

\subsection{Secondary Eclipses}

AC~Cnc also exhibits secondary eclipses.  Table~\ref{HalfWidths} also
lists the widths and ingress/egress phases for the secondary eclipses.
Figure~\ref{SecEcl} shows the eclipse widths versus orbit number for
pointings 1 and 2.  As one might expect, there is considerable scatter
in the widths stemming from the scatter in the ingress and egress
phases.  The two data sets look very similar, however.  We do not see
any orbit-dependent behavior related to the bursts.  It is difficult
to interpret the secondary eclipse behavior given that we do not know
the exact source of the eclipsed light.

\subsection{Summary}

As a summary of this section on the light curves, we note the
following in these plots:

\begin{itemize}
  \item the primary eclipses of both data sets are completely centered
    on phase 0 to a high degree.

  \item with the exception of the stunted burst eclipses, the eclipse
    ingresses and egresses of essentially all of the orbits occur at
    the same phases.  Note in particular that this behavior occurs
    even when the out-of-eclipse phases are relatively spread in
    counts (e.g., eclipses 1:41-50, 1:71-80);

  \item essentially {\it none} of the light curves reveal a
    significant pre-eclipse bright spot, especially before and during
    the stunted bursts;

  \item the secondary eclipses are more variable than the primary
    eclipses as one would expect given the nature of the eclipsed and
    eclipsing sources.

\end{itemize}

\section{Stunted Outbursts: Properties}

Stunted outbursts were first described by \cite{Honeycutt1998} based
on ${\sim}$6.5 years of data covering old novae and novalike systems.
The bursts likely represent the actual state of affairs of CVs that
have been described as in `continuous outburst.'\footnote{That phrase
  appears to have first been used by E. Guinan for an IUE proposal
  \citep{Guinan81}.}  Honeycutt et al. noted that these systems
revealed `repetitive ${\sim}$0.6 mag bursts, sometimes accompanied by
${\sim}$0.6 mag dips.'  They reported detecting stunted bursts in
about ${\sim}$20\% of the monitored systems.  More recently,
\cite{Mason2016} note that the stunted burst behavior resembles the
anomalous Z Cam behavior described by \cite{Simonsen2011}\footnote{As
  an editorial comment by the authors, we do not object to the
  connection, but if all CVs showing stunted bursts are called
  anomalous Z Cam CVs, the sheer numbers of objects calls into
  question the use of the adjective `anomalous.'  For this reason, in
  this paper we retain `stunted' to describe the behavior.}.

The outbursts of AC~Cnc visible in the {\it K2} data set stand out for
two reasons: first, they rise above the variability of the light curve
with significant width relative to other variations.  The mean light
level is ${\approx}$27-30K counts; the first burst rises to
${\sim}$35K counts and the second to ${\sim}$38K counts; the third
burst rises to at least 38K counts.

Second, the three bursts are delineated by the {\it bottoms} of the
primary eclipses (green lines in Figure~\ref{CompleteLC}): the eclipse
minima lie at an essentially constant 12.6K counts, to a high
precision, {\it except} during the bursts, when the minima in the
bursts only drops to ${\sim}$13.5K counts.  That comment is not to
suggest that the eclipse depth changes, but that ${\approx}0.1$ mag of
the extra light is not eclipsed and that this only appears to occur in
the bursts.  Measuring from a mean peak to the bottom of the eclipse,
bursts 1 and 2 cover 1.05 and 1.14 mag respectively, while the `bumps'
at orbits ${\sim}$1:60 and ${\sim}$1:85 are 0.95 and 0.98 mag,
respectively.  While burst 3 is incomplete, we still see the eclipse
bottoms increase to ${\sim}$13-14K counts at roughly the burst peak.

Fitting bursts 1 and 2 for FWHM and amplitude, we obtain FWHM values
of $3.4{\pm}0.2$ and $7.6{\pm}0.4$ days ($= 11.3{\pm}0.66$ and
$25.3{\pm}1.33$ orbits, respectively) and amplitudes of
${\approx}$0.42 and 0.36 mags for AC~Cnc.  From ground observations,
\cite{Honeycutt1998} measured an amplitude range of ${\approx}0.4$ to
${\approx}0.75$ mag and FWHM values from ${\approx}$4-7.5 days.  The
values for the two {\it K2} bursts fall within the envelope of ground
observations of all stunted bursts (Figure~\ref{stunted}).  The {\it
  K2} AC~Cnc points are shorter in duration than the mean of all
AC~Cnc ground observations, most likely because the ground data lack
the time resolution necessary to define the FWHM as cleanly as is
possible in the {\it K2} data.  As burst 3 is incomplete, we do not
include it.  We note, however, that the FWHM and amplitude are at
least similar to the values for burst 2.

The two different stunted outburst widths (${\sim}$3.4 and ${\sim}$7.6
days) seen in this {\it K2} data are suggestive of AC Cnc having two
characteristic durations, which we call `normal' and `extended',
respectively.  The statistics are admitedly quite poor but the idea is
supported by the behavior seen in dwarf novae.  In most well-observed
dwarf nova, the distribution of outburst widths is bimodal, with a
pair of well-defined peaks separated by (typically) 4 to 8 days.
\cite{Hack1993} compiled histograms of a number of dwarf novae
showing this effect.  Examples include X Leo \cite{Saw1982}, U
Gem \cite{Isles1976}, Z Cam \cite{Petit1961}, EM Cyg \cite{Brady1977}
and \cite{Szkody1984}, and SS Cyg \cite{Martel1961}.  The phenomenon
was modeled in SS Cyg by \cite{Cannizzo1993} using the accretion disk
limit cycle model for the outburst mechanism.  If the suggestion of a
bi-modal distibution of outburst widths is born out for AC Cnc, then
the outbursts in dwarf novae and the stunted outbursts in novalike CVs
may be related in interesting ways.  To address this possibility, 
we would need multiple stunted bursts obtained with sufficient
precision to discern whether a difference in burst width actually
exists.  On the basis of the measured {\it K2} widths, the bursts
differ, but in comparison to stunted bursts measured from the ground
(e.g., Figure~\ref{stunted}), we can not infer a difference.  It
remains an intriguing possibility to be resolved with future
(robotic?) observations.

\section{Discussion}

Uninterrupted observations of CVs allow us to see an accreting system
in its `orbit-to-orbit', or `ordinary', life.  With a sufficiently
broad range of CVs observed in that manner, disk modelers would have a
clear set of data to test not only `snap shot' physics but also
the evolution of a model.  The observations presented here of AC~Cnc
demonstrate that potential.  There are only a handful of additional
eclipsing CVs in the {\it Kepler} and {\it K2} datasets, all of which
are dwarf novae: the SU~UMa DN V477 Lyr \citep{Ramsay2012}; DN KIS
J192748.53+444724.5 \citep{Scaringi2013,Littlefair2014}; SU~UMa DN
CRTS J035905.9+175034 \citep{Littlefield2018}), and the C18 data set
of SU~UMa DN SDSS J081256.85+191157.8 (Schlegel \& Honeycutt 2019, in
preparation).  

That the bursts present in the {\it K2} data match observations
obtained from the ground confirms that the {\it K2} outbursts may be
identified as stunted bursts.  In contrast to the ground observations,
in which the bursts are usually defined by a handful of data points,
{\it K2} reveals the bursts in exquisite detail.  The single dip
following burst 2 also matches the properties of the dips described in
\cite{Honeycutt1998}.

As noted previously, the primary eclipses are essentially constant in
width, except for small changes during the peaks of the bursts.  This
implies that extra light spills out from behind the secondary during
primary eclipse.  Given the inferred inclination of the system
(75.6$\pm$0.7, \cite{Thoroughgood2004}), enhanced accretion disk
emission or a wind associated with a stunted burst could easily
explain the difference.

In contrast, the secondary eclipses demonstrate significant
variability.  If we adopt 22.5K counts as a very approximate `mean'
for data set 1, then there are excursions of +0.41 mag and -0.27 from
that value.  These variations are not present in the background, so
can not be attributed to other objects contaminating the {\it K2}
field-of-view.

The secondary eclipse behavior essentially tracks the overall light
curve (q.v., Figure~\ref{CompleteLC}), suggesting that mass transfer
explains the variations.  Yet the AC Cnc {\it K2} data also suggest
that mass transfer variations are not the explanation of stunted
bursts: we do not see enhanced brightening at the stream impact
point prior to the stunted bursts. 

In a typical CV model, we would attribute the variations to one or
more of three sources: the L$_1$-facing hemisphere of the secondary,
mass transfer stream + impact point, or accretion disk.  Any one or
all of these sources could contribute to the AC~Cnc secondary eclipse
variations, leaving us in a similar position as that of
\cite{Robertson2018}.  In that paper, the stunted bursts the authors
observed in UU Aqr confidently led them to conclude that mass transfer
via a blobby stream explained the stunted burst behavior.  But they
were equally acquainted with other systems strongly suggesting
accretion disk instabilities as explanations for stunted bursts.  The
slight difference in inclination of UU Aqr (78$^{\circ}{\pm}2^{\circ}$
\citep{Baptista1994}) vs AC Cnc ($75^{\circ}.6{\pm}0.7$
\citep{Thoroughgood2004}) is too small to have a large impact on the
interpretation of the data.

One must, however, also keep in mind that the eclipsing body for the
secondary eclipse is the accretion disk and associated structures.
Portions of the eclipsing body must be optically thick to produce a
$\approx$0.25 mag deep secondary eclipse.  If the optical depth of the
accretion disk changes, however, due to variable accretion, then the
depths of the secondary eclipses will also vary.  The lack of
pre-eclipse bright spots for AC Cnc appears to eliminate the stream
and impact point as sources of the variations, leaving the facing
hemisphere of the secondary, optical depth changes of the disk, or
enhanced structure in the disk as possible explanations.  For a
${\approx}$0.25 mag change, an enhancement in the number density or in
a small change in the disk structure, e.g., an enhancement in the disk
rim, by 10-20\% for number density and path length could account for
the secondary eclipse variations.  To quantify this more concretely
will require detailed fitting of each eclipse curve.  Such a study is
planned but lies outside the scope of the current paper.

We are also left with an additional intriguing question: at exactly
what point does an enhancement in the disk's light become a stunted
outburst?  Our question really boils down to the physics of the disk:
can we really suggest that stunted burst 1 differs significantly from
the three increases in flux centered on orbit numbers ${\sim}1:60$,
${\sim}1:85$, and ${\sim}1:125$?  One can argue that stunted burst 1
has a faster rise time and that property defines a burst.  But the
slope leading into the rise in stunted burst 2 (orbit numbers
${\sim}1:150-160$) is very similar to the slope leading into the peak
of the non-burst around orbit numbers ${\sim}1:50-55$.  What has
occurred to lead to a burst, stunted or not, in one case and not the
other?  The bursts are clearly set off by the primary eclipse minima,
illustrating that extra light is present, but what has changed?  Based
on the data set in hand, we do not have an answer.

\section{Summary}

We described two pointings of the novalike CV AC~Cnc using the {\it
  K2} satellite during Campaigns 5 and 18.  The data cover nearly
100\% of a continous 1-minute sequence of observations over 74 days
for data set 1 and about 50 days for data set 2.  The eclipse timing
defines an ephemeris statistically consistent with prior definitions
but provides {\it at best} marginal evidence for a period change --
directly countering a claim in the published literature.  The light
curves reveal two complete, and one partially observed, stunted
outbursts.  The properties of the stunted bursts match those obtained
from ground observations of novalike CVs.  The {\it K2} data fill in
the behavior of the stunted bursts, showing a continuously varying
evolution of the disk.

\vspace{0.1in}
\facility{Facility: Kepler2}

\section{ORCIDs}

EMS: 0000-0002-4162-8190; RKH: none

\acknowledgements

We thank the anonymous referee for comments that improved this paper.
This paper includes data collected by the Kepler mission during its
revamped K2 mission -- specifically, Campaigns 5 and 18. Funding for
the Kepler mission is provided by the NASA Science Mission
Directorate.  This research also made use of NASA's Astrophysics Data
System.  This research also make use of the software package PyKE
\citep{Still2012} distributed by the NASA {\it Kepler} Guest Support
office.  This work was supported by Kepler2 Grant GO3-4104Z and the
Vaughan Family Professorship.

\setcounter{figure}{2}
\begin{figure}
  \caption{(b) The complete light curve for data set 2 showing a stunted
    burst cut off by the fuel pump `hiccup.'
    A data gap exists at orbits 46 and 47.  See the caption to (a) for
    complete details.}
\scalebox{0.5}{\rotatebox{-90}{\includegraphics{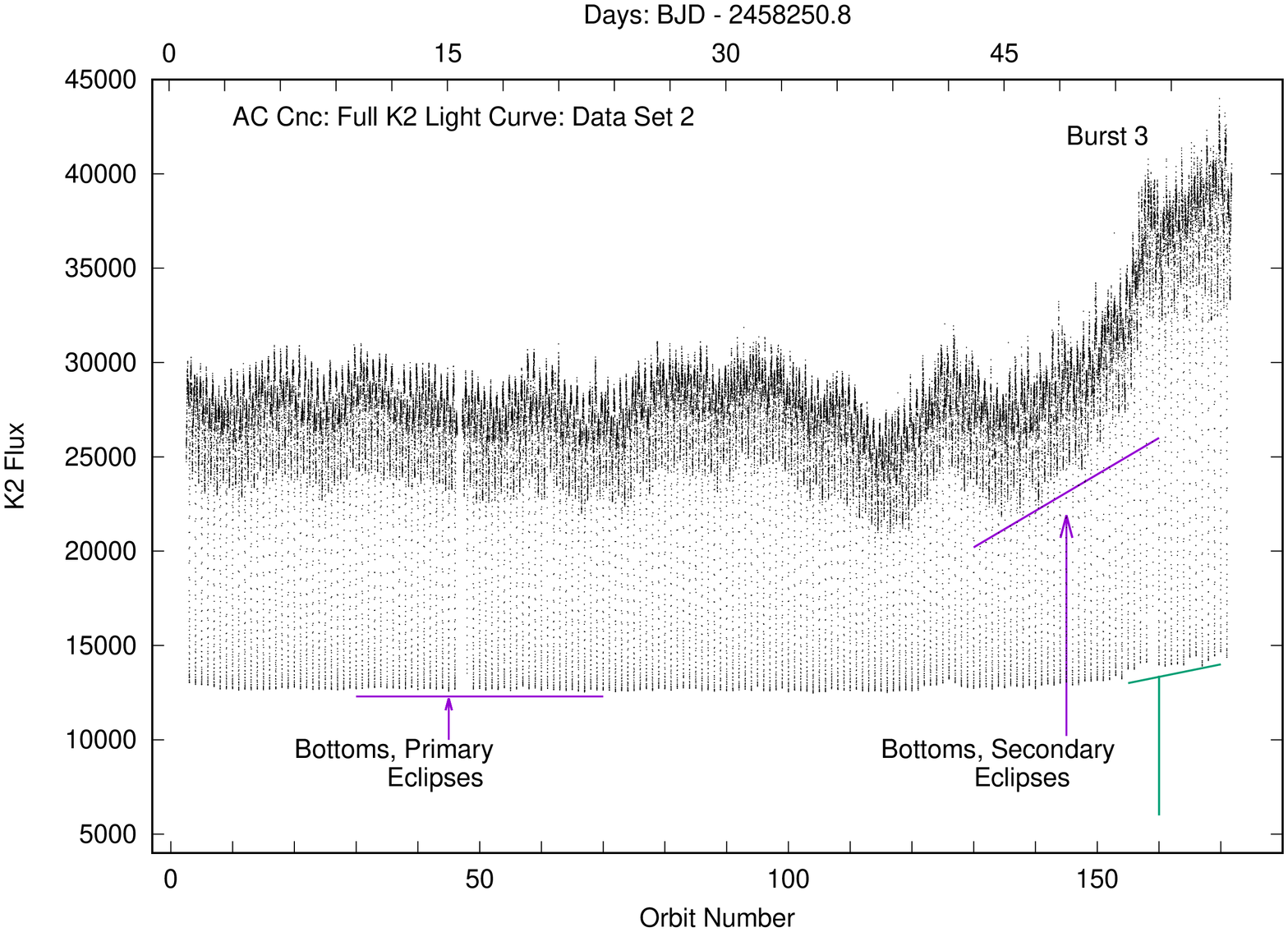}}}
\end{figure}  

\begin{figure}
  \caption{Primary eclipse half-width at half-maximum vs orbit
    number. The top plot shows the half width; the bottom shows the
    starting and ending phases.  Each set of points also shows the
    median (blue line) and mean (gold line) values -- for the starting
    phases, both median and mean are nearly identical; the half-width
    values are slightly separated: the median
    ${\sim}0.0952{\pm}0.0009$ vs the mean ${\sim}0.0932{\pm}0.0006$.
    The red rectangles indicate the orbits of bursts 1 (left) and 2 (right).
    (a) Data Set 1}
  \label{EclipWidth}
  \scalebox{0.5}{\rotatebox{-90}{\includegraphics{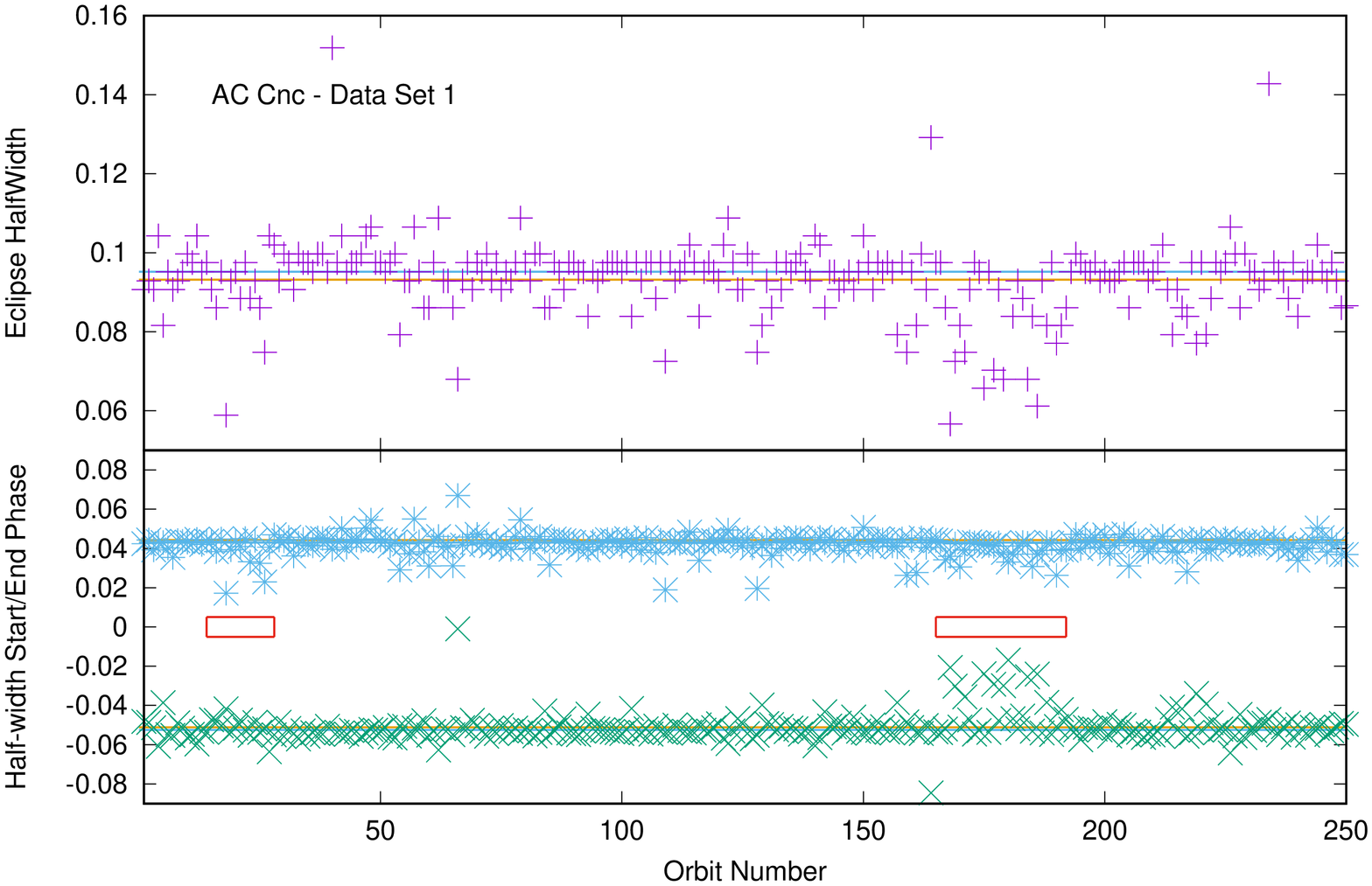}}}
\end{figure}

\setcounter{figure}{3}
\begin{figure} 
  \caption{(b) Primary eclipse half-width for data set 2.  The median
    value is $0.0943{\pm}0.0122$ vs. the mean value = $0.0929{\pm}0.00094$.
    The red rectangle indicates the orbits of burst 3.}
  \scalebox{0.5}{\rotatebox{-90}{\includegraphics{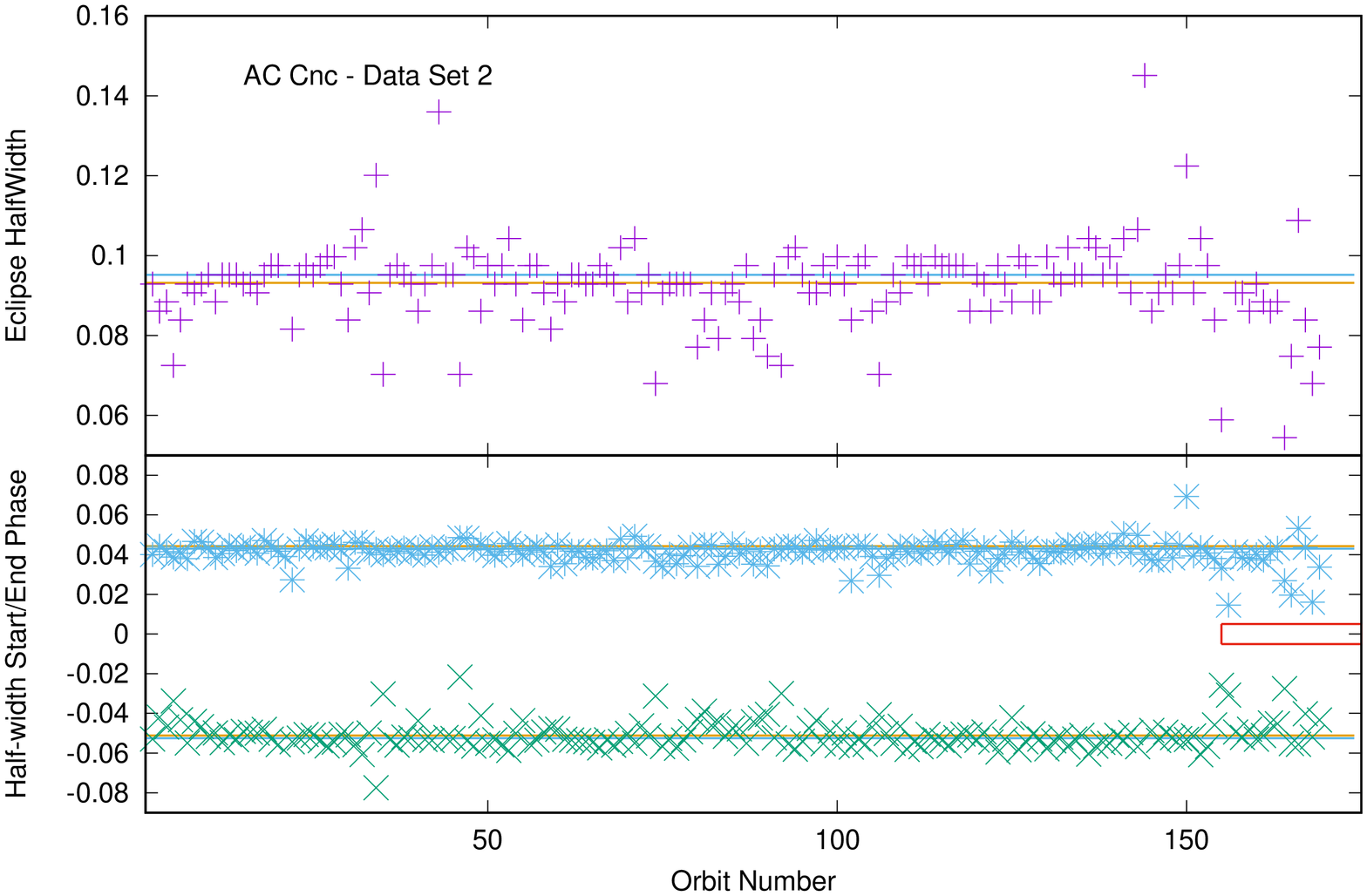}}}
\end{figure}

\begin{figure}
  \caption{Orbital light curves for two selected portions of the light
    curve of Data Set 1 (Figure~\ref{CompleteLC}(a)) where the
    selections are marked by colored boxes: (left) orbits
    ${\sim}$100.5 to 103.5 and (right) orbits ${\sim}$171.5 to 174.5.
    A short data gap typical of cosmic ray hits occurs just prior to
    eclipse at 168.7.}
  \label{ExpLC}
\scalebox{0.35}{\rotatebox{-90}{\includegraphics{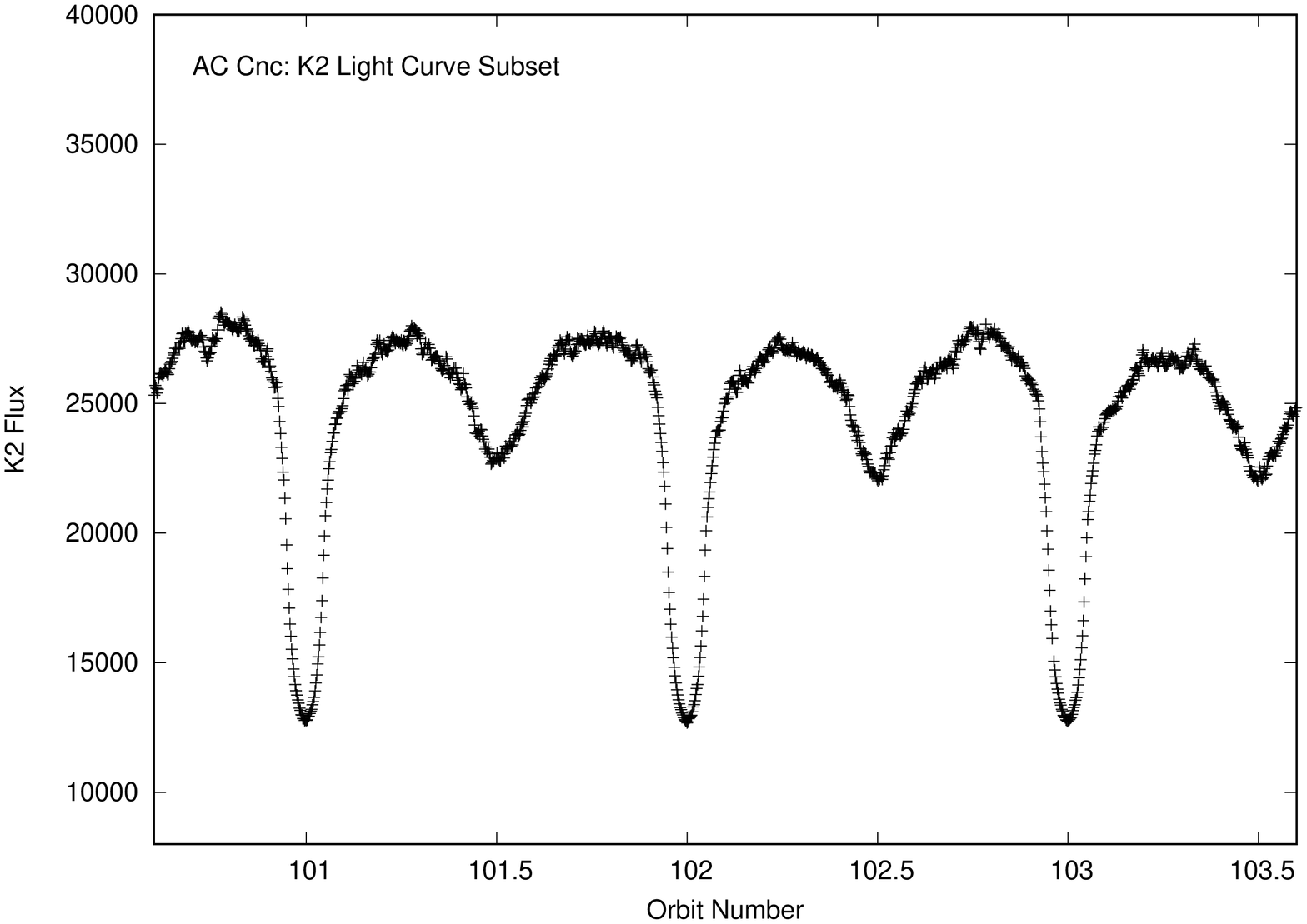}}}
\scalebox{0.35}{\rotatebox{-90}{\includegraphics{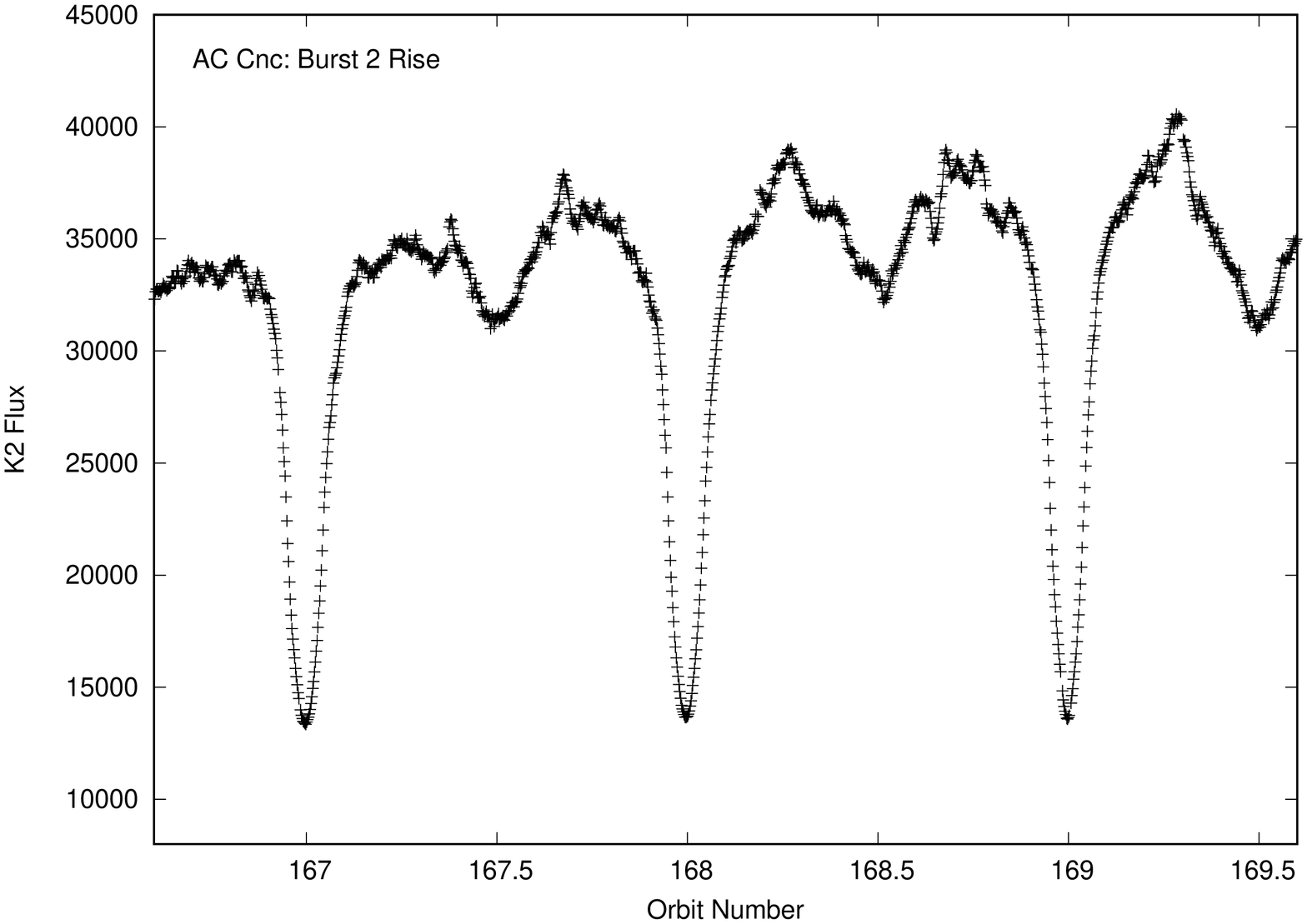}}}
\end{figure}

\begin{figure}
  \caption{The fully-phased light curve of AC~Cnc using a different
    color for each orbit.  The primary and secondary eclipses are
    readily apparent as is pseudo `two-state' light levels, `pseudo'
    because of the essentially continuous nature of the variations.
    Given this behavior, sampling the light curve at frequencies
    typical of ground-based observatories could easily lead to an
    inference of `state changes'.  (a) data set 1}
  \label{PhzdLC}
\scalebox{0.4}{\rotatebox{-90}{\includegraphics{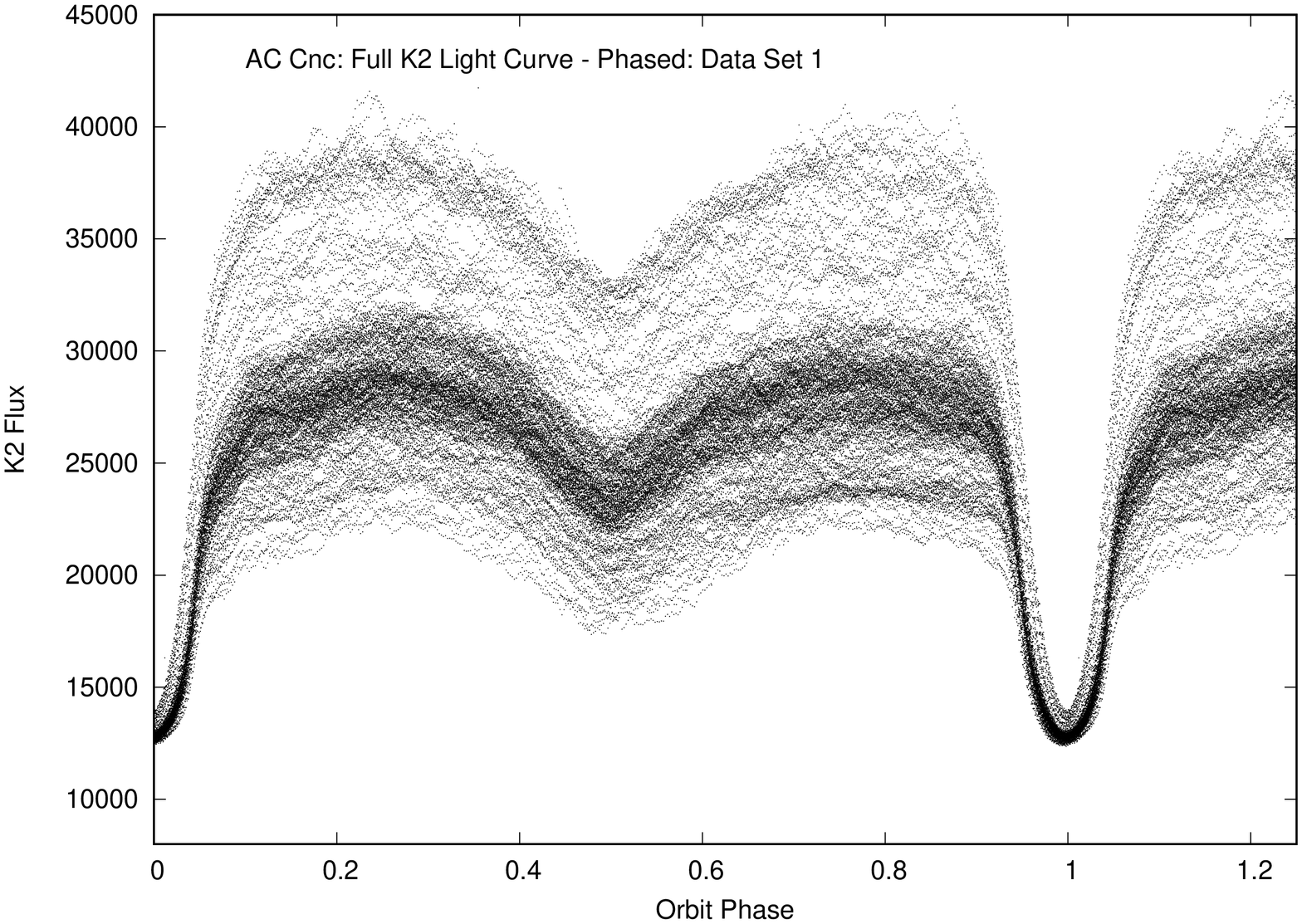}}}
\end{figure}

\setcounter{figure}{5}
\begin{figure}
  \caption{ (b) The fully-phased light curve for data set 2.  The cutoff
    of observations prevented catching the decline of the stunted burst.
    That makes the phased light curve resemble a state change.  See the
    text for a longer discussion.}
\scalebox{0.4}{\rotatebox{-90}{\includegraphics{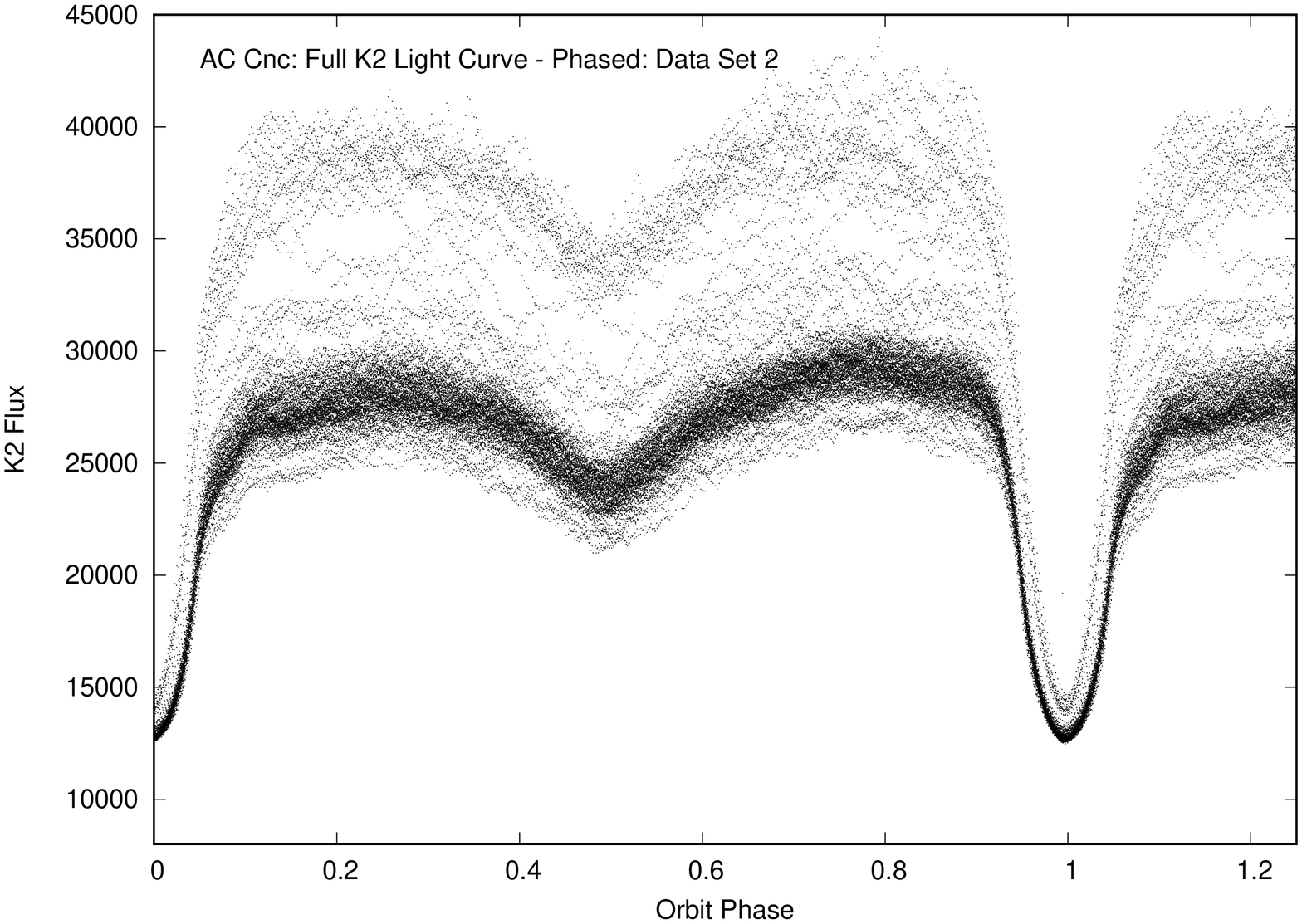}}}
\end{figure}

\begin{figure}
\caption{The complete set of light curves broken into 10-orbit blocks
  (the numbers inside each plot-let). The colors assigned to each light
  curve cycle through the same order for each plot-let, starting with
  black.  The color order is most readily visible in the 11-20
  plot-let (where the light curves are most affected by the stunted
  burst).  (a) data set 1, orbits 1-150.}
\label{plotlets}
\scalebox{0.6}{\rotatebox{-90}{\includegraphics{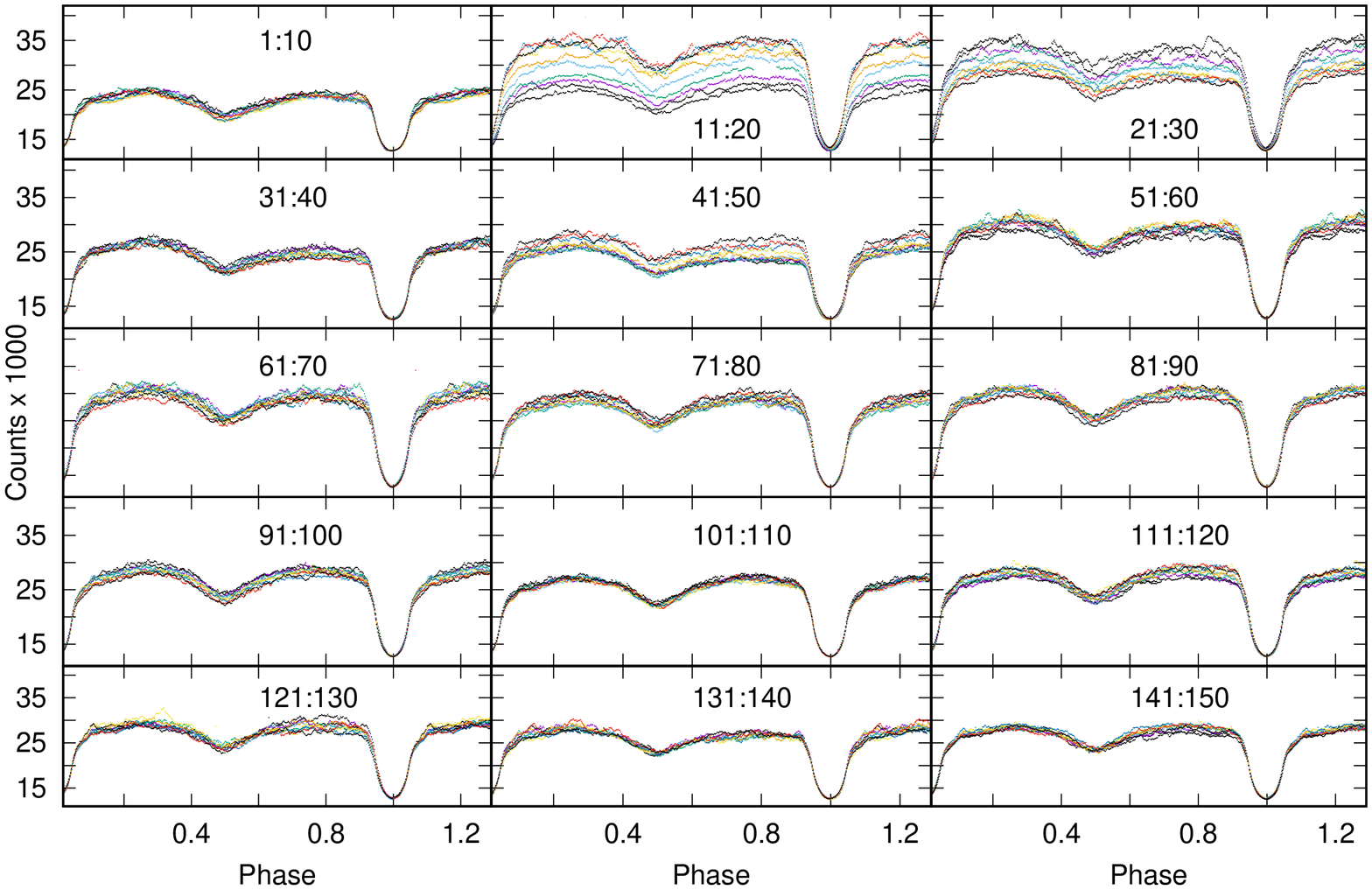}}}
\end{figure}

%\clearpage

\setcounter{figure}{6}

\begin{figure}
\caption{The complete set of light curves broken into 10-orbit blocks
  (the numbers inside each plot-let) (b) data set 1, orbits 151-246.}
%\label{plotlets}
\scalebox{0.6}{\rotatebox{-90}{\includegraphics{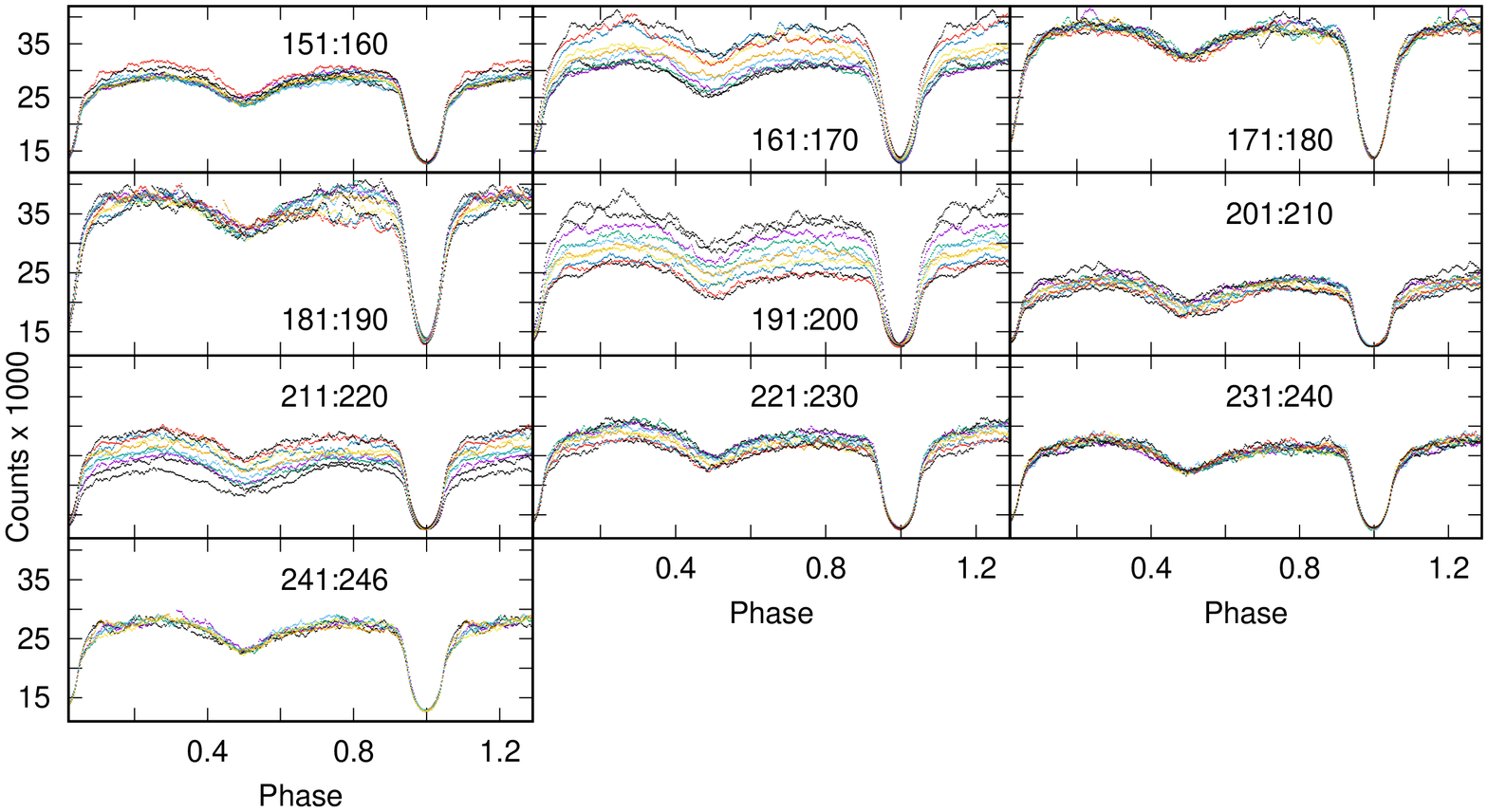}}}
\end{figure}

\setcounter{figure}{6}

\begin{figure}
\caption{The complete set of light curves broken into 10-orbit blocks
  (the numbers inside each plot-let) (c) data set 2, orbits 1-120.}
%\label{plotlets}
\scalebox{0.6}{\rotatebox{-90}{\includegraphics{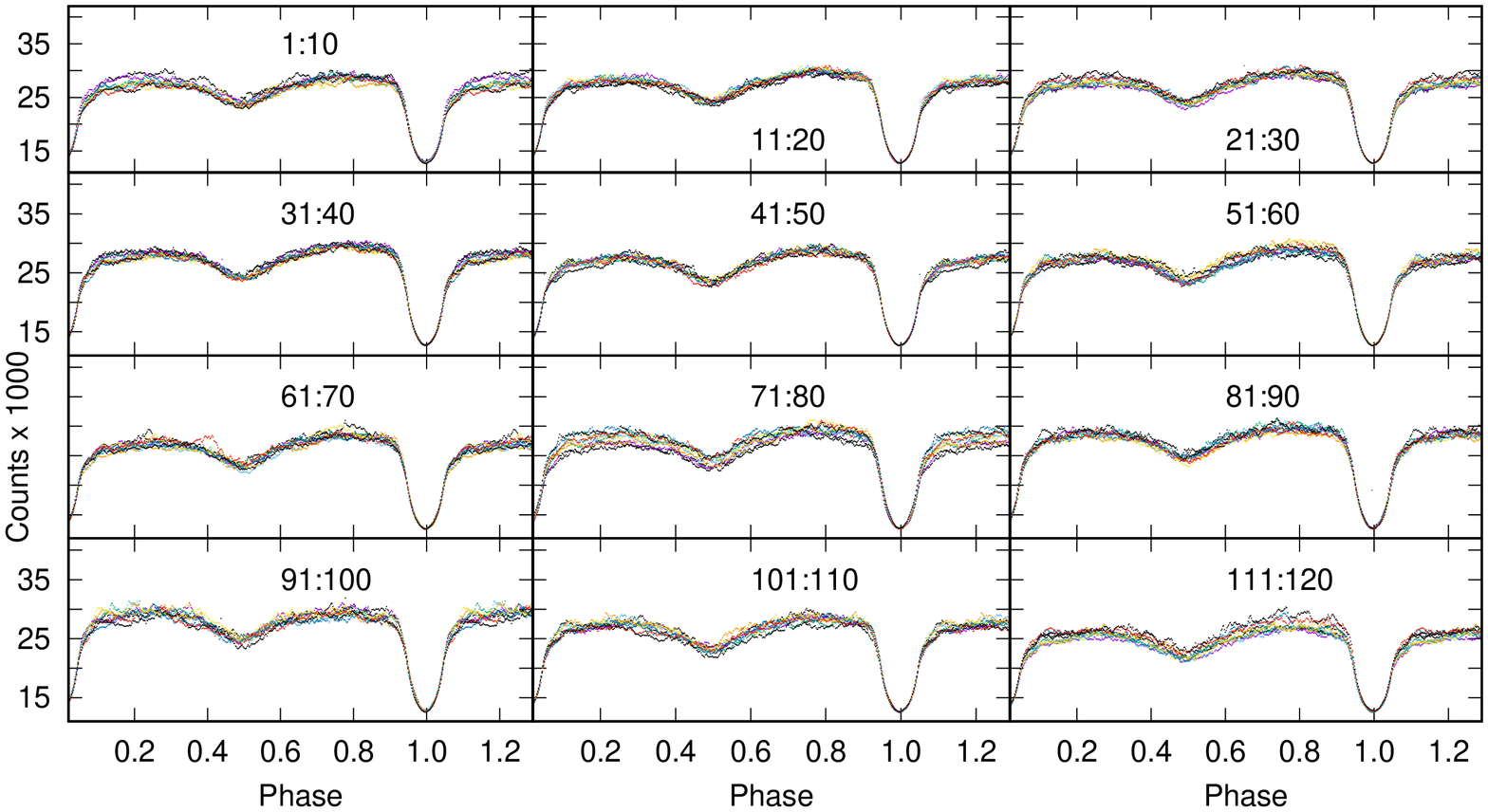}}}
\end{figure}

\setcounter{figure}{6}

\begin{figure}
\caption{The complete set of light curves broken into 10-orbit blocks
  (the numbers inside each plot-let) (b) data set 2, orbits 121-171.}
%\label{plotlets}
\scalebox{0.6}{\rotatebox{-90}{\includegraphics{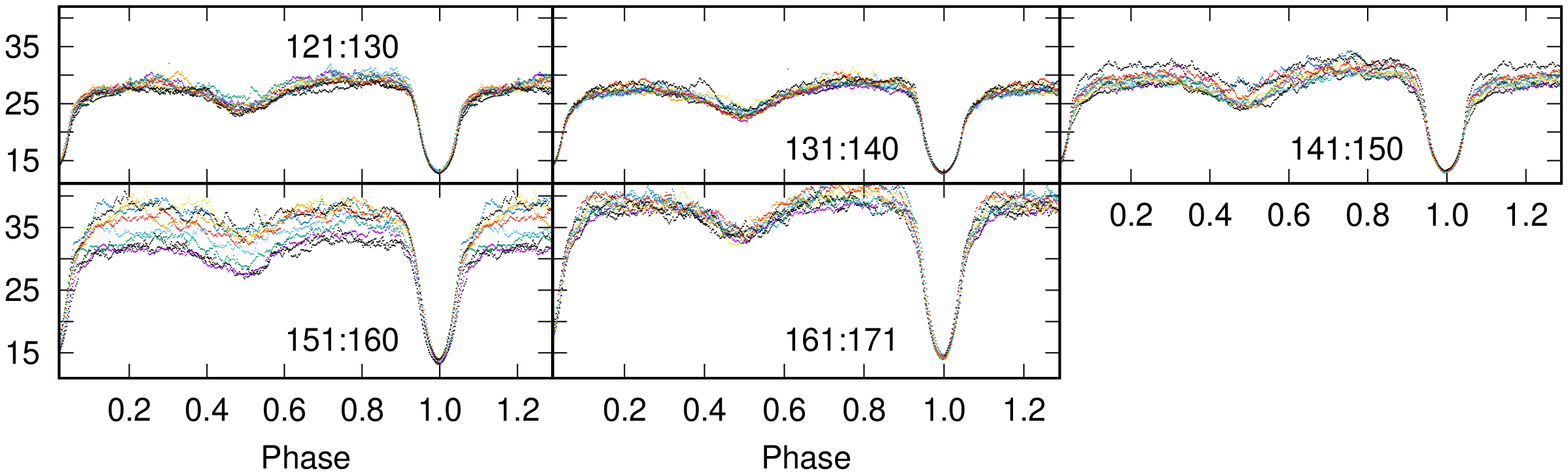}}}
\end{figure}

\begin{figure}
  \caption{Secondary eclipse widths for the two data sets.  (a) Data set 1.}
  \label{SecEcl}
  \scalebox{0.39}{\rotatebox{-90}{\includegraphics{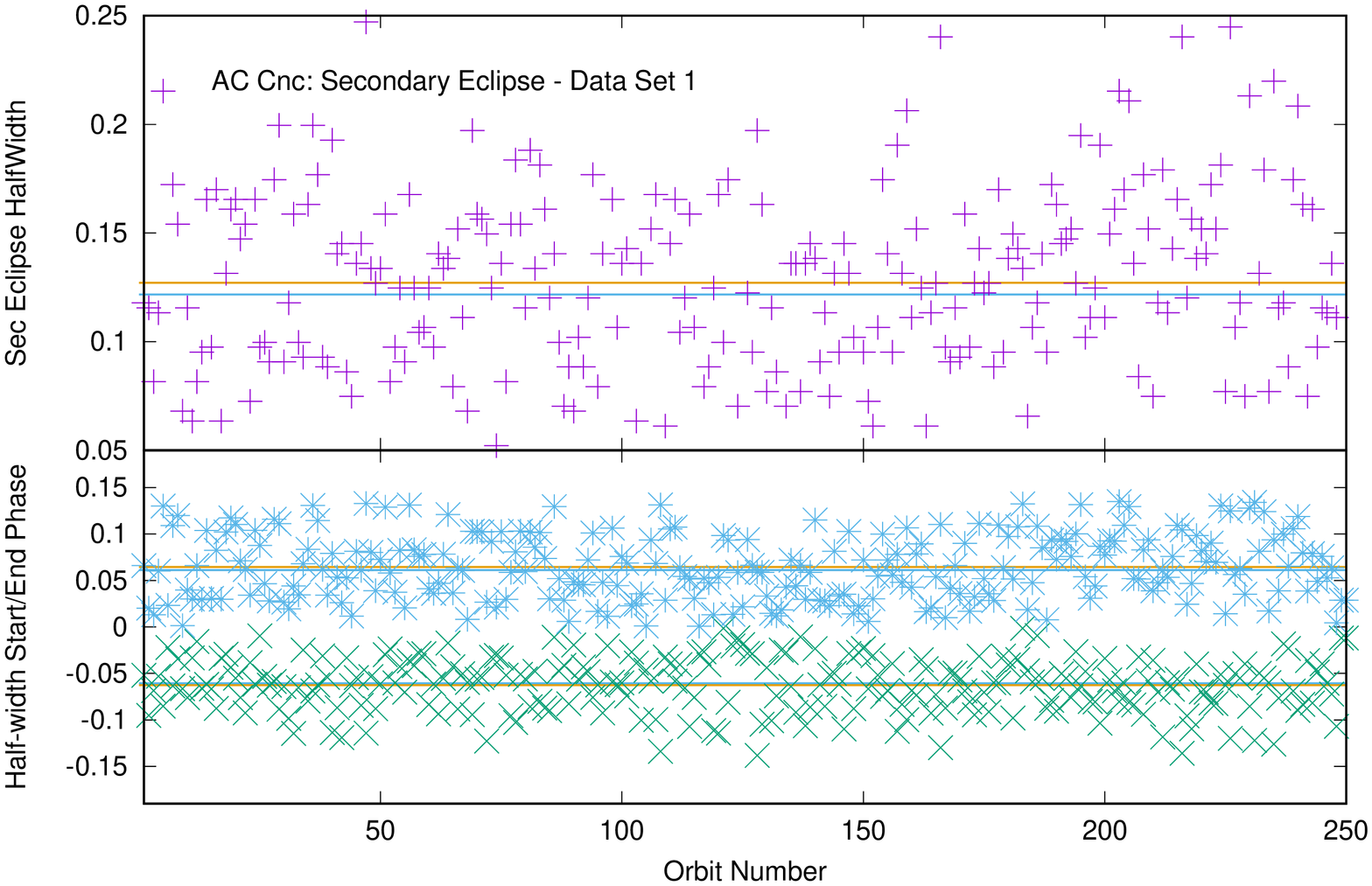}}}
\end{figure}

\setcounter{figure}{7}

\begin{figure}
  \caption{Secondary eclipse widths: (b) Data set 2.}
%  \label{SecEcl}
  \scalebox{0.39}{\rotatebox{-90}{\includegraphics{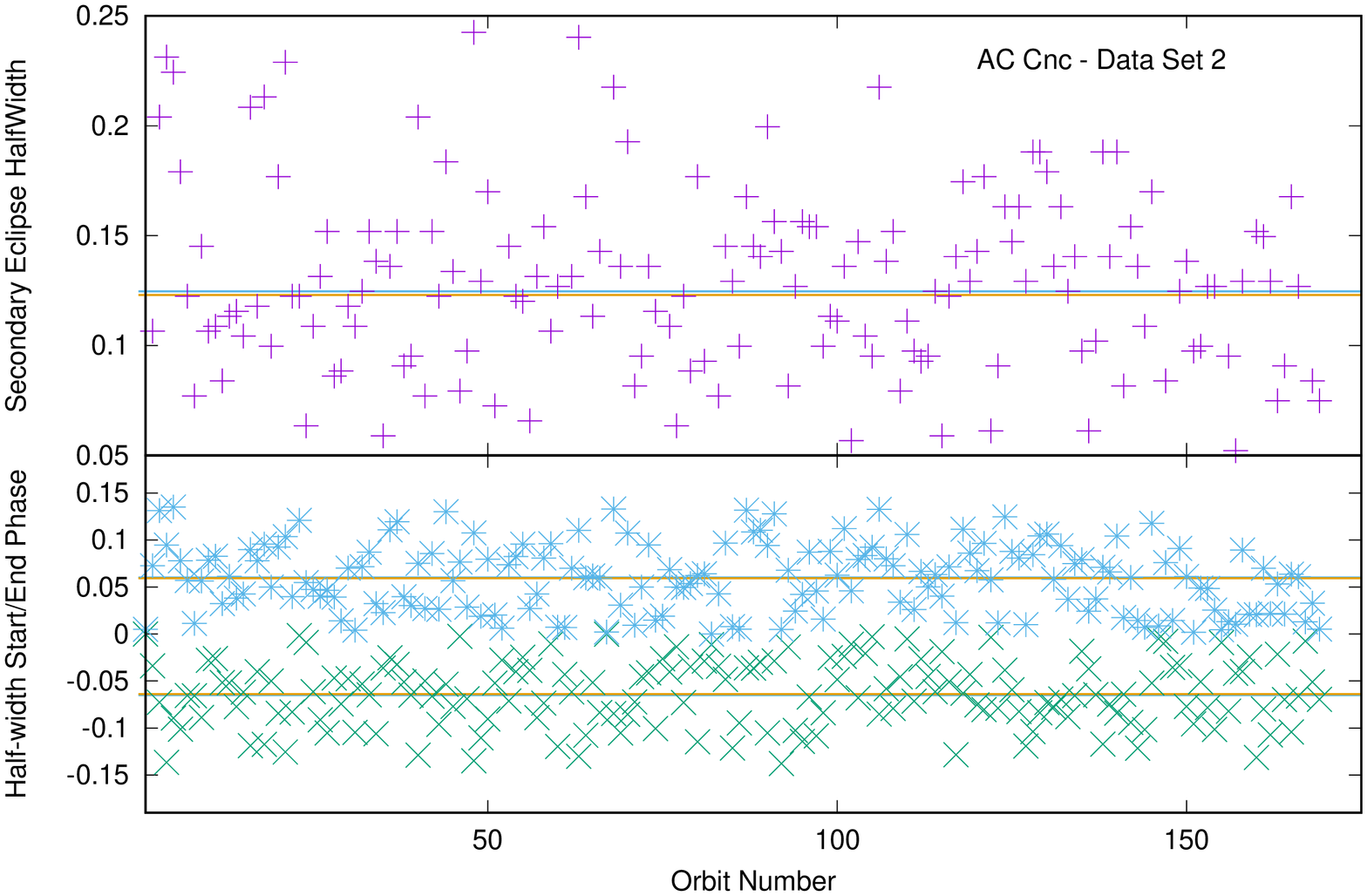}}}
\end{figure}

\begin{figure}
\caption{A comparison of the amplitude and FWHM of stunted outbursts
of the {\it K2} outbursts of AC Cnc during C05 with a number of cataclysmic
variables, including ground (grd) observations of AC Cnc.
Uncertainties are suppressed for visibility but are ${\sim}$0.5 day
(FWHM) and 0.05 mag in amplitude.}
\label{stunted}
\scalebox{0.4}{\rotatebox{-90}{\includegraphics{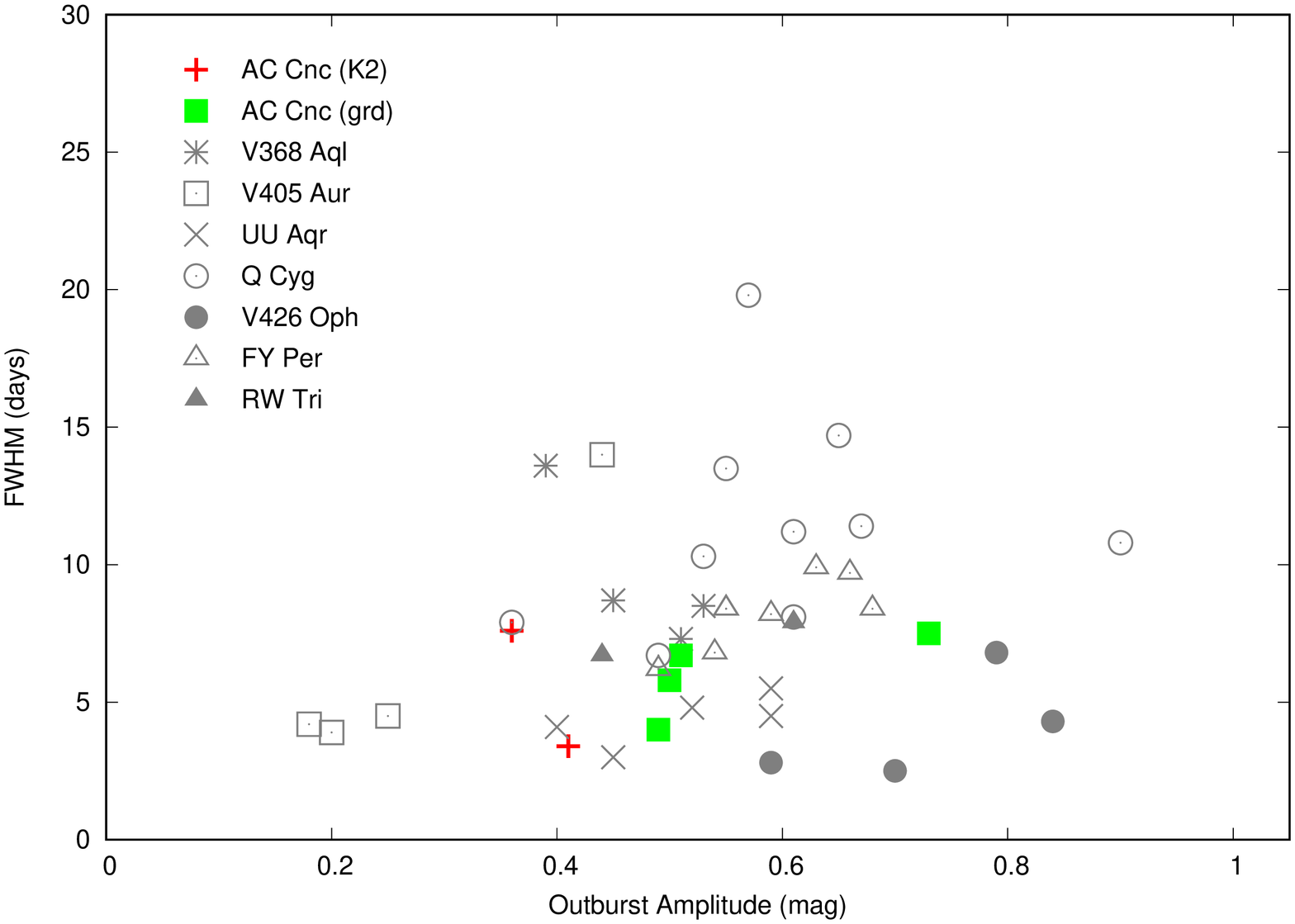}}}
\end{figure}


\begin{thebibliography}{}

\bibitem[Baptista et al.(1994)]{Baptista1994} Baptista, R., Steiner, J. E.,
  \& Cieslinski, D. 1994, ApJ, 433, 332

\bibitem[Brady \& Herczeg(1977)]{Brady1977} Brady, R.~A., \&
  Herczeg, T.~J.\ 1977, \pasp, 89, 71

\bibitem[Cannizzo(1993)]{Cannizzo1993} Cannizzo, J.~K.\ 1993,
  \apj, 419, 318

\bibitem[Dhillon et al.(2013)]{Dhillon2013} Dhillon, V. S., Smith,
  D. A., \& Marsh, T. R. 2013, MNRAS, 428, 3559
  
\bibitem[Dmitrienko(1995)]{Dmitrienko1995} Dmitrienko, E. S. 1995,
  Astr. Lett., 21, 171

\bibitem[Dmitrienko(1991)]{Dmitrienko1991} Dmitrienko, E. S. 1991, IzKry, 83, 131

\bibitem[Downes(1982)]{Downes1982} Downes, R. A. 1982, PASP, 94, 950

\bibitem[Eastman et al.(2010)]{Eastman2010} Eastman, J., Siverd, R.,
  Gaudi, B. S. 2010, PASP, 122, 935

\bibitem[Gies et al.(2013)]{Gies2013} Gies, D. R., Guo, Z., Howell,
  S. B., Still, M. D., Boyajian, T. S., Hoekstra, A. J., Jek, K. J.,
  LaCourse, D., Winarski, T. 2013, ApJ, 775, 64

\bibitem[Guinan(1981)]{Guinan81} Guinan, E. F. 1981, IUE Porposal ID \#CVDEG

\bibitem[Hack et al.(1993)]{Hack1993} Hack, M., Ladous, C., Jordan, S.~D., et al.
  1993, NASA Special Publication, 507

\bibitem[Honeycutt(2001)]{Honeycutt2001} Honeycutt, R. K. 2001, PASP, 113, 473

\bibitem[Honeycutt et al.(1998)]{Honeycutt1998} Honeycutt, R. K.,
  Robertson, J. W., \& Turner, G. W. 1998, AJ, 115, 2527

\bibitem[Howell et al.(2014)]{Howell2014} Howell, S. B., Sobeck, C.,
  Haas, M., Still, M., Barclay, T., Mullaly, F., Troeltzsch, J.,
  Aigrain, S., Bryson, S. T., Caldwell, D., Chaplin, W. J., Cochran,
  W. D., Huber, D., Marcy, G. W., Miglio, A., Najita, J. R., Smith, M.,
  Twicken, J. D., \& Fortney, J. J. 2014, PASP, 126, 938

\bibitem[Isles(1976)]{Isles1976} Isles, J.~E.\ 1976, Journal of the
  British Astronomical Association, 86, 327

\bibitem[Kepler(2013)]{Kepler2014} Kepler Data Characteristics Handbook
  (KSCI-19040-004, dated 31 May 2013)
  
\bibitem[Kinemuchi et al.(2012)]{Kinemuchi2012} Kinemuchi, K., Fanelli, M.,
  Pepper, J., Still, M., \& Howell, S. B. 2012, PASP, 124, 963

\bibitem[K\"ording et al.(2011)]{Kording2011} K\"ording, E. G.;
  Knigge, C.; Tzioumis, T.; \& Fender, R. 2011, MNRAS, 418, L129

\bibitem[Kreiner(2004)]{Kreiner2004} Kreiner, J. M. 2004, ActaAstron,
  54, 207

\bibitem[Kurochkin \& Shugarov(1980)]{Kurochkin1980} Kurochkin, N. E. \&
  Shugarov, S. Yu 1980, Astr. Tsirk. 1114

\bibitem[Kurochkin \& Shugarov(1981)]{Kurochkin1981} Kurochkin, N. E. \&
  Shugarov, S. Yu 1981, Astr. Tsirk. 1154

\bibitem[Littlefair et al.(2014)]{Littlefair2014} Littlefair, S. P.,
  Dhillon, V. S., G\"ansicke, B. T., Bours, M. C. P., Copperwheat, C. M.,
  \& Marsh, T. R. 2014, MNRAS, 443, 718

\bibitem[Littlefield et al.(2018)]{Littlefield2018} Littlefield, C.,
  Garnavich, P., Kennedy, M., Szkody, P., \& Dai, Z. 2018, AJ, 155, 232

\bibitem[Martel(1961)]{Martel1961} Martel, L.\ 1961, Annales d'Astrophysique, 24, 267

\bibitem[Mason \& Howell(2016)]{Mason2016} Mason, E. \& Howell,
  S. B. 2016, A\&A, 589, 106

\bibitem[Patterson(2012)]{Patterson2012} Patterson, J. 2012, JAVSO,
  40, 240

\bibitem[Petit(1961)]{Petit1961} Petit, M. 1961, Asiago Contrib 119, 31

\bibitem[Provencal et al.(2014)]{Provencal2014} Provencal, J. L.,
 Shipman, H. L., Montgomery, M. H., \& WET Team 2014, Contr. of
 Astron. Obs. Skalnaté Pleso, 43, 524

\bibitem[Qian et al.(2007)]{Qian2007} Qian, S.-B.; Dai, Z.-B.; He,
 J. J.; Yuan, J. Z.; Xiang, F. Y.; \& Zejda, M. 2006, A\&A, 466, 589

\bibitem[Ramsay et al.(2016)]{Ramsay2016} Ramsay, G., Hakala, P.,
 Wood, M. A., Howell, S. B., Smale, A., Still, M. D., \& Barclay,
 T. 2016, MNRAS, 455, 2772

\bibitem[Ramsay et al.(2012)]{Ramsay2012} Ramsay, G., Canizzo, J. K.,
 Howell, S. B., Wood, M. A., Still, M., Barclay, T., \& Smale, A.,
 2012, MNRAS, 425, 1479
   
\bibitem[Robertson et al.(2018)]{Robertson2018} Robertson, J. W.;
  Honeycutt, R. K.; Henden, A. A.; \& Campbell, R. T. 2018, AJ, 151, 61

\bibitem[Saw(1982)]{Saw1982} Saw, D.~R.~B.\ 1982, Journal of the
  British Astronomical Association, 92, 220

\bibitem[Scaringi et al.(2013)]{Scaringi2013} Scaringi, S., Groot,
  P. J., \& Still, M. 2013, MNRAS, 435, 68
   
\bibitem[Schlegel et al.(1984)]{Schlegel1984} Schlegel, E. M.;
  Honeycutt, R. K.; Kaitchuck, R. H. 1984, ApJ, 280, 235

\bibitem[Shugarov(1981)]{Shugarov1981} Shugarov, S. Yu. 1981, SovAstr, 25,
  332

\bibitem[Simonsen(2011)]{Simonsen2011} Simonsen, M. 2011, JAAVSO, 39, 66
  
\bibitem[Still \& Barclay(2012)]{Still2012} Still, M. \& Barclay,
  T. 2012, Astrophysics Code Library, code ascl:1208.004

\bibitem[Szkody \& Mattei(1984)]{Szkody1984} Szkody, P., \& Mattei, J.~A.\
  1984, \pasp, 96, 988

\bibitem[Thoroughgood et al.(2004)]{Thoroughgood2004} Thoroughgood,
  T. D., Dhillon, V. S., Watson, C.A., Buckley, D. A. H., Steeghs, D.,
  \& Stevenson, M. J.  2004, MNRAS, 353, 1135

\bibitem[Torbett \& Campbell(1987)]{Torbett1987} Torbett, M. V. \&
  Campbell, B. 1987, ApJ, 318, L29

\bibitem[Warren et al.(2006)]{Warren2006} Warren, S. R., Shafter,
  A. W., Reed, J. K. 2006, PASP, 118, 1373
  
\bibitem[Yamasaki, Okazaki, \& Kitamura(1983)]{Yamasaki1983} Yamasaki,
A; Okazaki, A; \& Kitamura, M. 1983, PASJ, 35, 423


\end{thebibliography}
\end{document}